\documentclass[journal,comsoc]{IEEEtran}
\usepackage[T1]{fontenc}

\usepackage{cite}

\ifCLASSINFOpdf
   \usepackage[pdftex]{graphicx}
\else
   \usepackage[dvips]{graphicx}
\fi
\usepackage{amsmath}
\usepackage[cmintegrals]{newtxmath}
\usepackage[linesnumbered,lined,ruled]{algorithm2e}

\usepackage[inline]{enumitem}
\usepackage{array}
\usepackage{tabularx}

\newenvironment{conditions*}
  {\par\vspace{\abovedisplayskip}\noindent
   \tabularx{\columnwidth}{>{$}l<{$} @{${}={}$} >{\raggedright\arraybackslash}X}}
  {\endtabularx\par\vspace{\belowdisplayskip}}

\ifCLASSOPTIONcompsoc
  \usepackage[caption=false,font=normalsize,labelfont=sf,textfont=sf]{subfig}
\else
  \usepackage[caption=false,font=footnotesize]{subfig}
\fi

\usepackage{float}
\usepackage{dblfloatfix}
\usepackage{url}

\usepackage{xcolor}

\makeatletter  
\def\@IEEEBIOhangwidth{2.6cm}    
\def\@IEEEBIOhangdepth{3cm}    
\makeatother

\begin{document}
%
\title{Profile-based Resource Allocation for Virtualized Network Functions}
%
%
%

\author{\IEEEauthorblockN{
Steven Van Rossem\IEEEauthorrefmark{1},
Wouter Tavernier\IEEEauthorrefmark{1},
Didier Colle\IEEEauthorrefmark{1},\\ 
Mario Pickavet\IEEEauthorrefmark{1} and
Piet Demeester\IEEEauthorrefmark{1}}
\IEEEauthorblockA{\IEEEauthorrefmark{1}Ghent University - imec, IDLab.\\
Email:  \{steven.vanrossem, wouter.tavernier, didier.colle, mario.pickavet, piet.demeester\}  @ugent.be}
}

\maketitle

\begin{abstract}
The virtualization of compute and network resources enables an unseen flexibility for deploying network services. A wide spectrum of emerging technologies allows an ever-growing range of orchestration possibilities in cloud-based environments. 
But in this context it remains challenging to rhyme dynamic cloud configurations with deterministic performance. 
The service operator must somehow map the performance specification in the Service Level Agreement (SLA) to an adequate resource allocation in the virtualized infrastructure. 
We propose the use of a VNF profile to alleviate this process.
This is illustrated by profiling the performance of four example network functions (a virtual router, switch, firewall and cache server) under varying workloads and resource configurations.
We then compare several methods to derive a model from the profiled datasets.
We select the most accurate method to further train a model which predicts the services' performance, in function of incoming workload and allocated resources.
Our presented method can offer the service operator a recommended resource allocation for the targeted service, in function of the targeted performance and maximum workload specified in the SLA.
This helps to deploy the softwarized service with an optimal amount of resources to meet the SLA requirements, thereby avoiding unnecessary scaling steps. 

\end{abstract}


%
\IEEEpeerreviewmaketitle

\section{Introduction}
%
%
%
%
\IEEEPARstart{T}{he} advancements in the domain of cloud
computing, Software Defined Networking (SDN) and Network
Function Virtualization (NFV) enable a unseen flexibility and programmability of both compute and network configurations.
By softwarizing network functions, we move away from dedicated hardware based, monolithic systems to a virtualized solution for offering telecom services.
The service is decomposed into multiple microservices which each get an allocated share of resources such as CPU time, memory access or network bandwidth.
Typical tasks involved in network services include packet forwarding, routing, inspection or any other form of network traffic processing.
Beyond the application layer, the deeper layers of the network traffic are checked or manipulated in a chained configuration.
This means that network traffic is sequentially steered through a, possibly lengthy, chain of processors such as routers, firewalls, load-balancers or proxy-servers.
In the NFV domain, the main aim is to provide softwarized solutions for each of those network functions, which can be deployed on commercial-of-the-shelf (COTS) servers. Ideally, equally high performance is expected compared to rigid, dedicated hardware middleboxes, but at a lower cost, higher flexibility regarding scaling, configuration and less prone to vendor and technology lock-in.

At deployment time of the network service, an estimation of the required capacity and related resource allocation needs to be made. The performance contract is given in the Service Level Agreement (SLA) and should be translated to the required resources.
In case of a hardware based middlebox, performance can be more easily guaranteed and specified, as this is a controlled and isolated environment. The internal processing is completely under control and validated by the middlebox vendor. Configuration settings are tested and specified in the vendor's test environment.
But in this case, the total resource reservation is not flexible and often resulting in an over-provisioned and expensive amount of rigid middleboxes, calculated to support the maximum expected workload. 
When using Virtual Network Functions (VNFs) instead, the resource reservation translates to the amount of virtualized compute and network resources needed to process the real-time workload, e.g. the number of  vCPUs, memory and bandwidth which must be reserved to support the current number of users. Over time, the amount of resources can be adjusted dynamically and more fine-grained.
However, characterizing or modelling the performance and required resources of such a VNF is not a straightforward task. The softwarized nature of VNFs implies a much larger space of possible hardware and software configurations, which can influence the resource usage and performance in many unexpected directions.

\begin{figure}[!b]
\centering
\includegraphics[clip=true, trim=0 220 0 0, width=\textwidth]{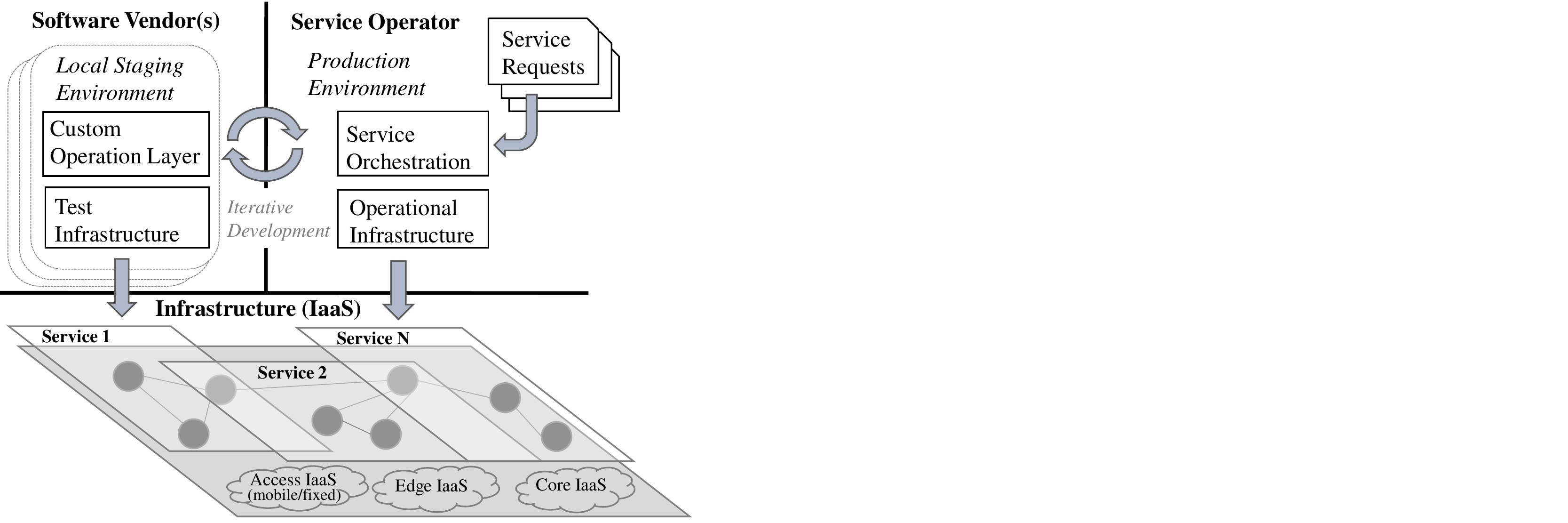}
\caption{A common IaaS environment helps to profile the VNF in a representative context and infrastructure.}
\label{fig_devops}
\end{figure}

\newpage
In Fig. \ref{fig_devops}, we can see how cloud-native Infrastructure-as-a-Service (IaaS) management enables new dynamics in using test and operational infrastructures. We have outlined such a platform architecture in \cite{ngpaas},  where  Service Operators can source VNFs from multiple Vendors to integrate them in their services. Building further on well-known DevOps methodologies, the vendor and operator can share the same IaaS environment to validate and operate VNFs. This helps to profile the VNF in a representative operational context, similar to where the operator would deploy it. 
In another previous publication \cite{SDKpaper} we had also advocated the practical use of a sandbox environment for validating VNFs prior to deployment in production and also adopted the idea to profile the VNF in this sandboxed environment.
Virtualized IaaS and DevOps methodologies create an ideal framework for automated VNF profiling, as several challenges exist:

\noindent
(i) It is impractical to exhaustively validate the performance in all possible
situations due to limited time, budget
and infrastructure availability. We must select a representative subset of infrastructure and workload configurations to profile the VNF on.
In a datacenter for example, it might be sufficient to profile a VNF on one node (using vertical scaling) and than extrapolate its performance when scaling out horizontally to similar hardware nodes.

\noindent
(ii) The operator must often consider the VNF as a
black box because the internal implementation is not
exposed by the vendor. Without a provided analytical performance model, the VNF performance needs to be characterized through testing. A black-box profiling approach has the benefit that VNF functionality is also validated with representative workloads. Every NFV use-case will have its own specific workloads, and it is hard to capture all VNF flexibility in an analytical model. We argue that profiling through testing can offer a more trusted approach compared to a theoretical model of the internal VNF workings, derived by source code analysis.

In the following sections we will give details of our measurements and the implementation of a VNF profiling method.
Next, in Section \ref{related}, we discuss other related research where we build further on. Then, in Section \ref{data_gathering}, the tested VNFs and used measurement setup are described. In Section \ref{analysis} we compare and select the best analysis methods to derive a model for the VNF performance. In the remaining Sections \ref{usecases}
and \ref{conclusion} we discuss practical use cases for a VNF profile.


\section{Related Work}
\label{related}
Unsurprisingly, the underlying server hardware characteristics have a deep impact on the performance. Parameters such as processor architecture, clock rate, size of the internal processor cache, memory latency,
bandwidth of inter-processor and peripheral buses, etc. have a strong impact on the performance of the specific application or VNF running on that server.
An extensive list of capabilities for bare metal and virtualized environments can be found in \cite{ETSIperf}. There it is also described how descriptor files can help to more strictly orchestrate VNF performance to specific hardware.
To accurately profile and reproduce the VNF performance, the service operator must be aware of the factors which can influence the service performance. 
Possible factors are listed in Table \ref{table_iaas} with references for measured results in the literature.
Since it is practically impossible to exhaustively profile all possible configurations, the service provider should be aware of the settings and control them as good as possible.
This is also illustrated in \cite{fang2018evaluating} where vswitch performance is very much depending on factors such as  traffic mix, scheduler settings and other deployment configuration options.

\begin{table}[!t]
\renewcommand{\arraystretch}{1.3}
\caption{Versatility of the infrastructure configuration in cloud environments.}
\label{table_iaas}
\centering
\begin{tabular}{|p{1.4cm}|p{6cm}|}
\hline
\textit{Situation} & \textit{Type of settings} \\
\hline
\hline
Orchestration & Instantiation latency caused by orchestration platform implementation. \cite{ventre2016performance} \\
\hline
\hline
Operating System \newline(Kernel space) 
\newline
+
\newline 
Hypervisor (Dom0) 
& CPU pinning.
\newline  Kernel-bypassing with network polling drivers (e.g. DPDK, netmap, FD.io).
\cite{pitaev2017multi}
\newline Resource scheduling optimization (vCPU time, memory, packet processing) \cite{kulkarni2017nfvnice} \cite{katsikas2017profiling}.
\newline Kernel network buffers and queue tweaks \cite{networktuning}.
\newline Hypervisor processing overhead (Xen) \cite{chen2015profiling}.
\\
\hline
\hline
Bare metal 
&  NIC (line rate, TCP processing offload, SR-IOV). 
\newline  CPU (clock speed, hyperthreading support).
\newline Memory (layered cache \cite{veitch2017cache}, RAM, DMA).
\newline Disk IO speed.
\newline Offload to specialized (accelerated) hardware e.g. GPU, FPGA \cite{duan2017separating}.
\\
\hline
\hline
Network 
& Congestion control, delay, routing protocols, 
\newline Degradation due to virtualization overhead \cite{shea2014deep}.\\
\hline
\end{tabular}
\end{table}

Different frameworks to obtain VNF profiles have been described before:
\cite{VNFgym} {\color{black} and \cite{nfvvital} describe an architecture to implement a VNF profiling framework where users can compare the
performance of VNFs in a controlled environment with multiple types of workload.
Similarly, the work in \cite{peuster2016understand} explains how an automated VNF profiling system was implemented, compatible with a DevOps approach.}
{\color{black}The automation of profiling measurements is further exemplified on chains of multiple VNFs in \cite{peuster2017profile}. The authors propose that a chain of VNFs should in fact be considered as a single entity for profiling. }  
These publications do however not provide an analysis method to model the profiled data for performance prediction.

An extensive overview of different statistical prediction methods for resource allocation is given in \cite{amiri2017survey}, however without quantitative results to compare the methods.
Validated methodologies to predict the VNF performance from earlier measurements are presented in \cite{vperfguard} and \cite{iglesias2017orca}. Curve fitting is used to model the relation between VNF performance metrics and input workload. The proposed methods monitor which resource and workload metrics are most correlated with a given KPI metric. The fitted relation is checked and adapted in real-time using a sliding window of the most recent samples. There is however no quantative comparison with other modelling methods, nor is discussed how initial resource allocation can be improved by profiling prior to deployment in production.

A different method is presented in \cite{lange2019discrete} and \cite{prados2017analytical} where the VNF performance is modelled using queuing theory. The monitored KPI metrics of the VNFs include e.g. buffer size, arrival rate and process rate. But only limited configurations are tested (with a fixed resource allocation) and it is not investigated if this method can be used to map SLA parameters to a recommended resource allocation.  
Other drawbacks of this method are that specialized probes are needed to monitor the different queue sizes in the VNF, which can be impossible for black-box/proprietary VNFs. Furthermore, as described in \cite{iglesias2017orca}, queue size or process rate can change dynamically when the VNF gets saturated, which is not captured in the proposed queueing models.

\section{Gathering Data for Performance Modelling}
\label{data_gathering}
A common knowledge from the machine learning domain is that a learning model is often only as good as the data used to train it.
Since our goal is to predict the performance of a VNF  as good as possible, care must be taken during the data gathering process that the performance of the VNF is  measured in a representative way. This is done in our measurement setup by isolating the resources used by the Device Under Test (DUT) and taking care that the traffic source/sink are not saturated during the measurements. During VNF profiling, we try to specify the performance of a VNF, within certain boundary conditions. To specify this, we categorize the monitored metrics into four types:
\begin{itemize}[leftmargin=*]
\item \textbf{Workload metrics} are used to quantify the amount of 'work' which is presented at the VNF's input. These reflect the user generated load (e.g. pps, packet size, requests/s, the variety in payload content or L2/3 header fields).
 
\item \textbf{Resource metrics} quantify the payable/physical/scalable hardware resources obtained from the IaaS provider and allocated to the VNF (e.g. vCPU, MEM, storage or network).
Also OS or hypervisor related metrics bound to the IaaS are considered (e.g. context switches or cache usage).

\item \textbf{Performance metrics} monitor the Key Performance Indicators (KPI). These are (often SLA-defined) measurements of the processed workload, thus taking  the output of the VNF into account  (e.g. delay, loss or throughput).

\item \textbf{\color{black}Context parameters} are one-shot IaaS  settings  (e.g. buffer length, scheduler algorithm, see Table \ref{table_iaas}) or VNF specific configurations (e.g. firewall rules or routing table length). This is part of the initialization and assumed fixed after deployment. In a IaaS context, many hardware settings are strictly limited or even completely shielded by the infrastructure provider. The VNFs are tested with one fixed context setting.

\end{itemize}

Categorizing the metrics like this, helps us to define which metrics should be monitored in the first place. After the measurements, we will need to train the VNF profile to predict the performance metrics from the resource allocation and workload metrics.
For each of the tested VNFs we will later specify the metrics more in detail.

\subsection{Measurement Setup}
\begin{figure}[!h]
\centering
\includegraphics[width=0.5\textwidth]{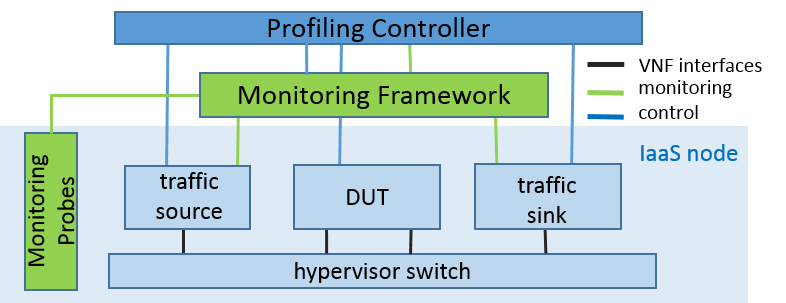}
\caption{The measurement setup used for profiling the VNFs.}
\label{setup}
\end{figure}

Figure \ref{setup} represents the different functional blocks used in the measurement setup. 
The DUT is the VNF which is being profiled, it gets an input and output interface.  The test traffic is routed through a hypervisor switch, from the traffic source, through the DUT, to the traffic sink.
The \textbf{Profiling Controller} iterates over every tested workload and records the monitored metrics for further analysis. A control interface must be foreseen in the DUT and traffic source/sink for initial configuration and to start/stop the workload.
The \textbf{Monitoring Framework} is configured by the Controller to gather metrics exported by a range of monitoring probes and exported by the traffic source/sink. We use Prometheus as framework. The used probes are cAdvisor, Prometheus Node Exporter and a custom tool to export Virtual Machine metrics gathered by KVM and libvirt.

The VNFs are running on a compute node with 2x 8core Intel E5-2650v2 (2.6GHz) CPU with Ubuntu 16.04. Linux Bridge is used as the hypervisor switch.
We do not change the default OS options (e.g. we leave hyperthreading enabled).
Depending on the virtualization of the VNF (container or Virtual Machine (VM)) we use the configuration options of Docker resp. KVM to isolate the CPU cores between the DUT and the traffic sink/source. {\color{black} }

\subsection{Measurement Strategy}
The monitoring capabilities of the platform where the VNFs are tested provide the base for our additional analysis. The general data gathering workflow works like this:
\begin{enumerate}[leftmargin=*]

\item We define the workload and resource configuration boundaries to deploy the DUT. These should be representative for the expected values possible in the production environment. Ideally the same type of IaaS node is used for profiling and in production. In between these boundaries we manually specify a number of intermediate values. Likewise, we also define the KPIs which quantify the DUT performance.

\item The metrics which represent the generated workload, allocated resources and VNF performance should be monitorable by the Monitoring Framework. The Profiling Controller instructs the Monitoring Framework to gather the required metrics.

\item The Profiling Controller configures the DUT and traffic source/sink with the specified resource allocation, and starts the workload. The Controller iterates over all combinations of the workloads and resource allocations defined in step 1. After each tested configuration is stabilized (see next subsection), the representative metric values are recorded and the next configuration is tested.

{\color{black}
\item The Monitoring Framework is configured to alert the Controller if either the traffic source/sink is overloaded (i.e. at their max cpu usage). In this case, the measured performance is marked as invalid, as it is bounded by the traffic VNFs and not by the profiled DUT. These measurements are not included in the profiled DUT model.}
\item When all measurements are completed, we have recorded the performance of the VNF in a large range of possible configurations. Each configuration depicts a certain input workload and resource allocation. This dataset is then analyzed further in section \ref{analysis}.

\end{enumerate}

\subsection{Assessing Measurement Stability}
\label{stability}
Each workload should only be generated long enough until a representative measurement can be taken. 
To assess the stability of an ongoing measurement, we derive a stability indicator which is based on the geometric mean of all monitored metrics $x_i$ together:

\begin{equation}\label{geom_mean}
\left(\prod _{i=1}^{n}x_{i}\right)^{\frac {1}{n}}={\sqrt[{n}]{x_{1}x_{2}\cdots x_{n}}} = \exp \left[{\frac {1}{n}}\sum _{i=1}^{n}\ln x_{i}\right]
\end{equation}

It can be seen in Eq. \ref{geom_mean} that the geometric mean converges to a stable value, if the sum of the logs of the metrics also becomes stable.
This has the advantage that metrics with different scales can be combined in the stability indicator. The sum of the logs will change proportionally to a relative change in any of the metrics (e.g. if one metric varies 5\%, the geometric mean will change proportionally, regardless of the scale of the metric). By taking the logs, we also limit the risk of calculation overflow.
When a workload is started, the measured metrics will stabilize after a certain ramp-up time. 
We assume that under a fixed workload, the resources and performance metrics $x_i$ will converge to a stable value, being constant with a certain Gaussian measurement noise. Under the Central Limit Theorem, the sum of the logs will then be approximately normal, so the mean and standard deviation should stabilize also.
The central tendency of the complete set of profiled metrics is therefore monitored by Eq. \ref{stability_metric}. Every second, we monitor $S$ in a moving window of the last 10sec. $\Delta S$ depicts the difference between consecutive windows of the $mean$ and $std$.

\begin{equation}
\label{stability_metric}
\Delta S = \begin{cases}
\Delta mean(\sum _{i=1}^{n}\ln x_{i})  <  \varepsilon_1  \\
\Delta std(\sum _{i=1}^{n}\ln x_{i})  <  \varepsilon_2
\end{cases}
\end{equation}

When $\Delta S$ stays below the thresholds for the last 5 sliding windows, we assume the measurement is stable and record the mean of the last 10sec of every monitored metric $x_i$. The thresholds $\varepsilon_{1,2}$ can be calibrated prior to the profiling test; under a stable workload, we can monitor the lower limits of $\Delta mean$ and $\Delta std$ on our setup. The stability indicator $\Delta S$ has been implemented in the monitoring framework itself and an event is fired to the Profiling Controller when stability is detected. By having only one single stability metric we decrease the monitoring overhead. This removes the need for the Profiling Controller to constantly poll all of the monitored metrics to check if the measurement has stabilized. If the metrics are not stable after 60sec, the workload is stopped and no measurements are recorded. In our setup, a stable measurement is detected after 30sec on average. This means it takes about 30sec to test a single configuration.

\begin{figure*}[!t]
\centering
\subfloat[OVS CPU usage]{\includegraphics[width=45mm]{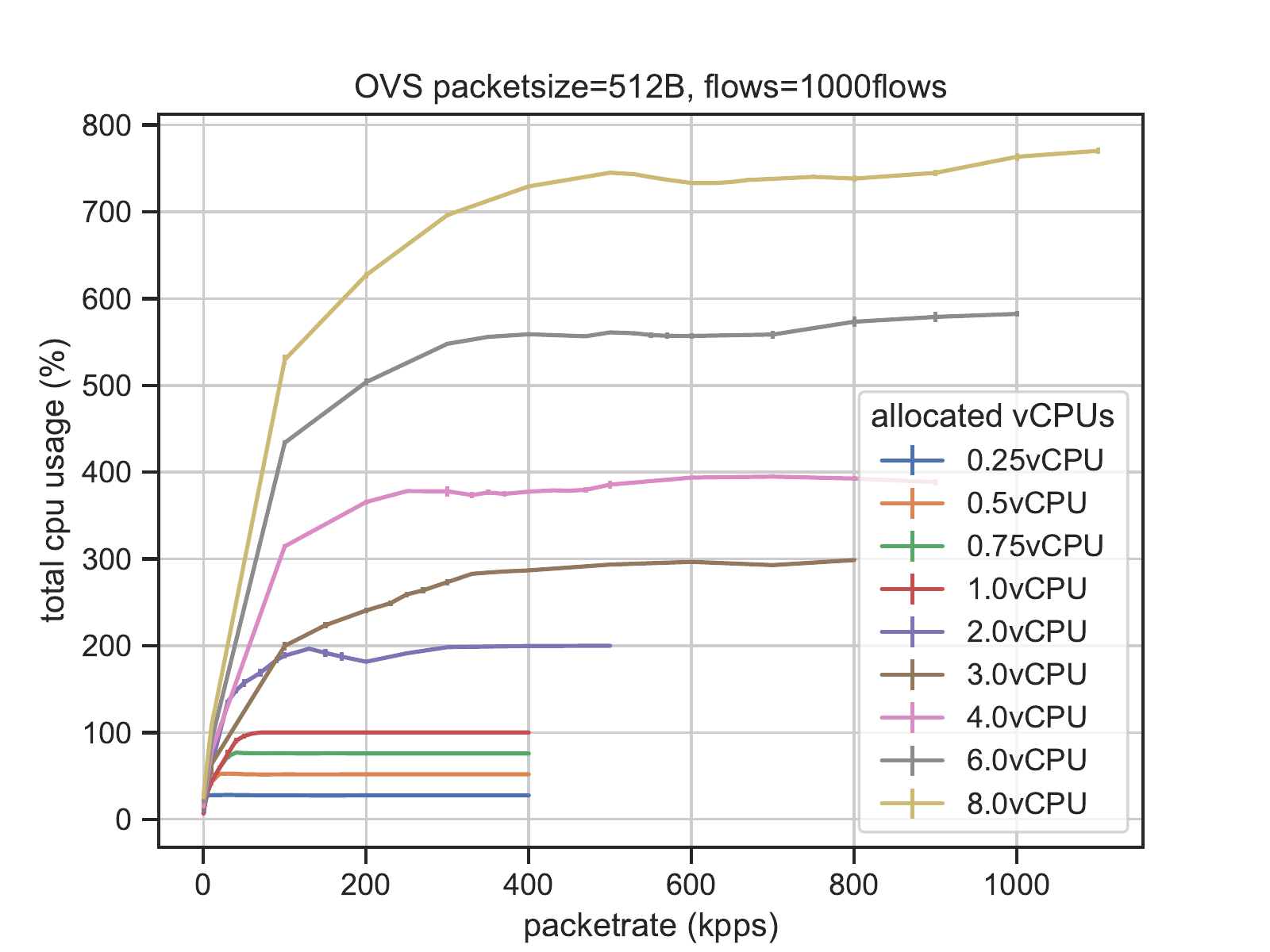}%
\label{ovscpu}}
\hfil
\subfloat[Router CPU usage]{\includegraphics[width=45mm]{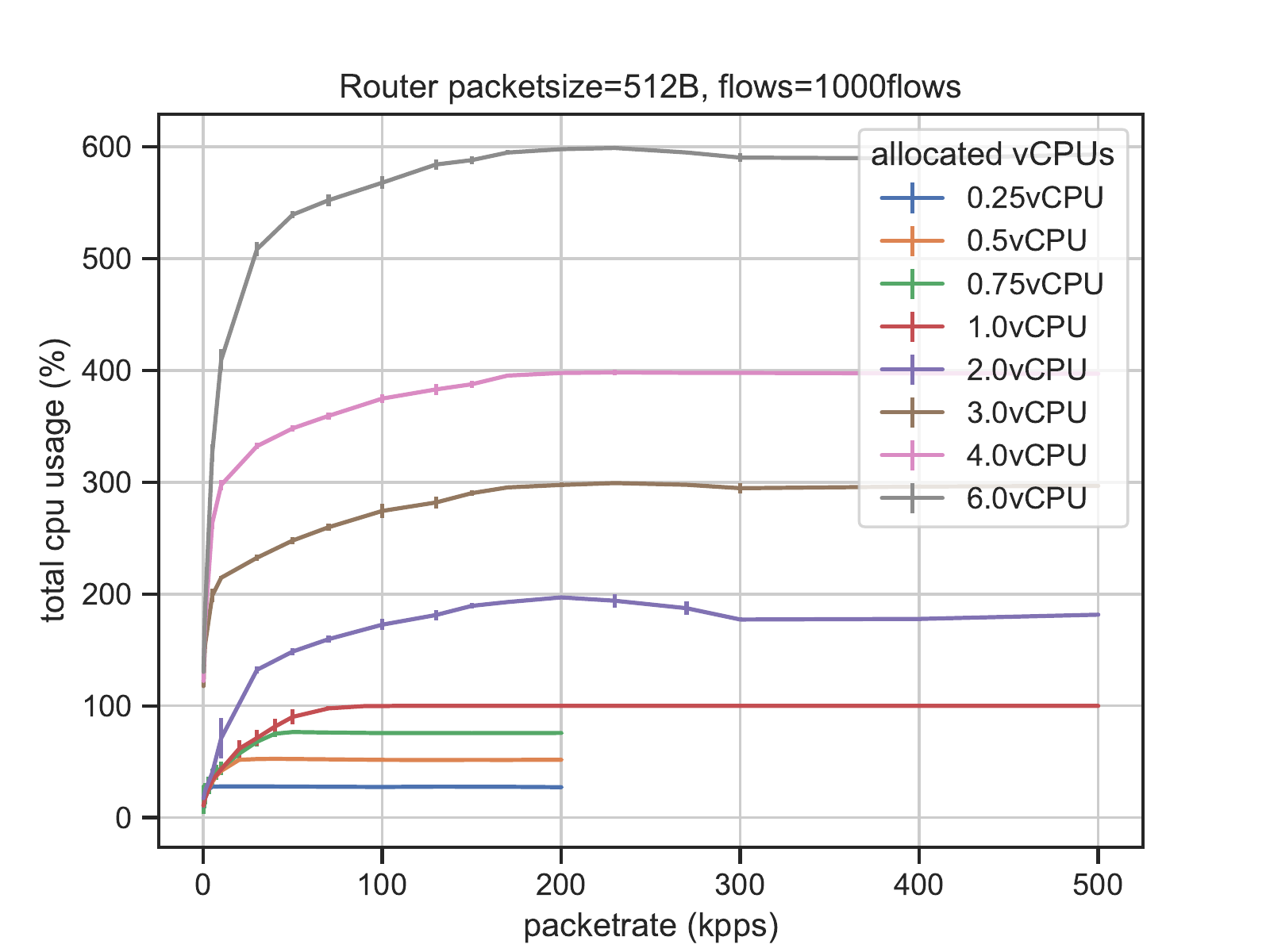}%
\label{routerecpu}}
\hfil
\subfloat[Firewall CPU usage]{\includegraphics[width=45mm]{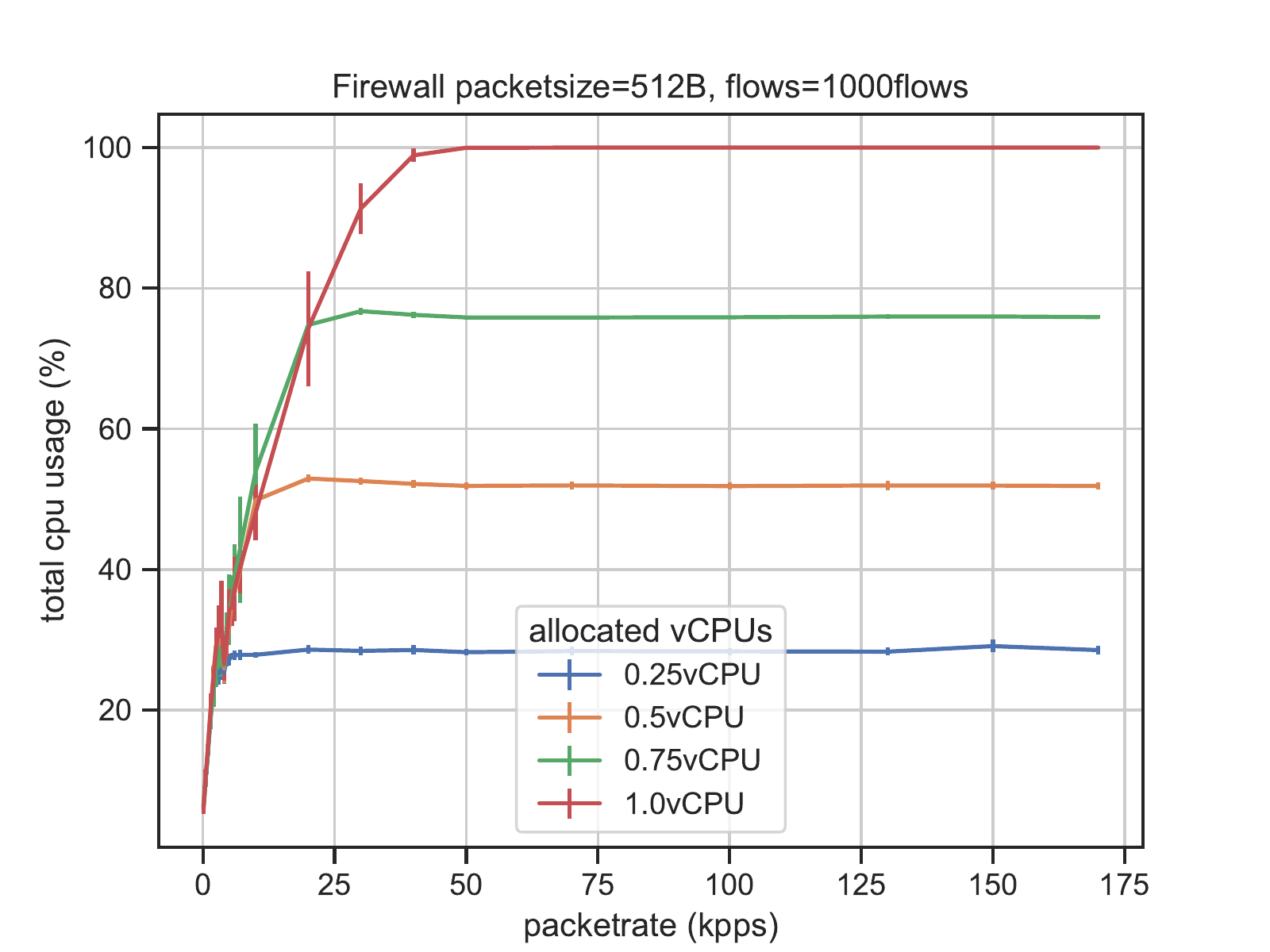}%
\label{fwcpu}}
\hfil
\subfloat[Cache CPU usage]{\includegraphics[width=45mm]{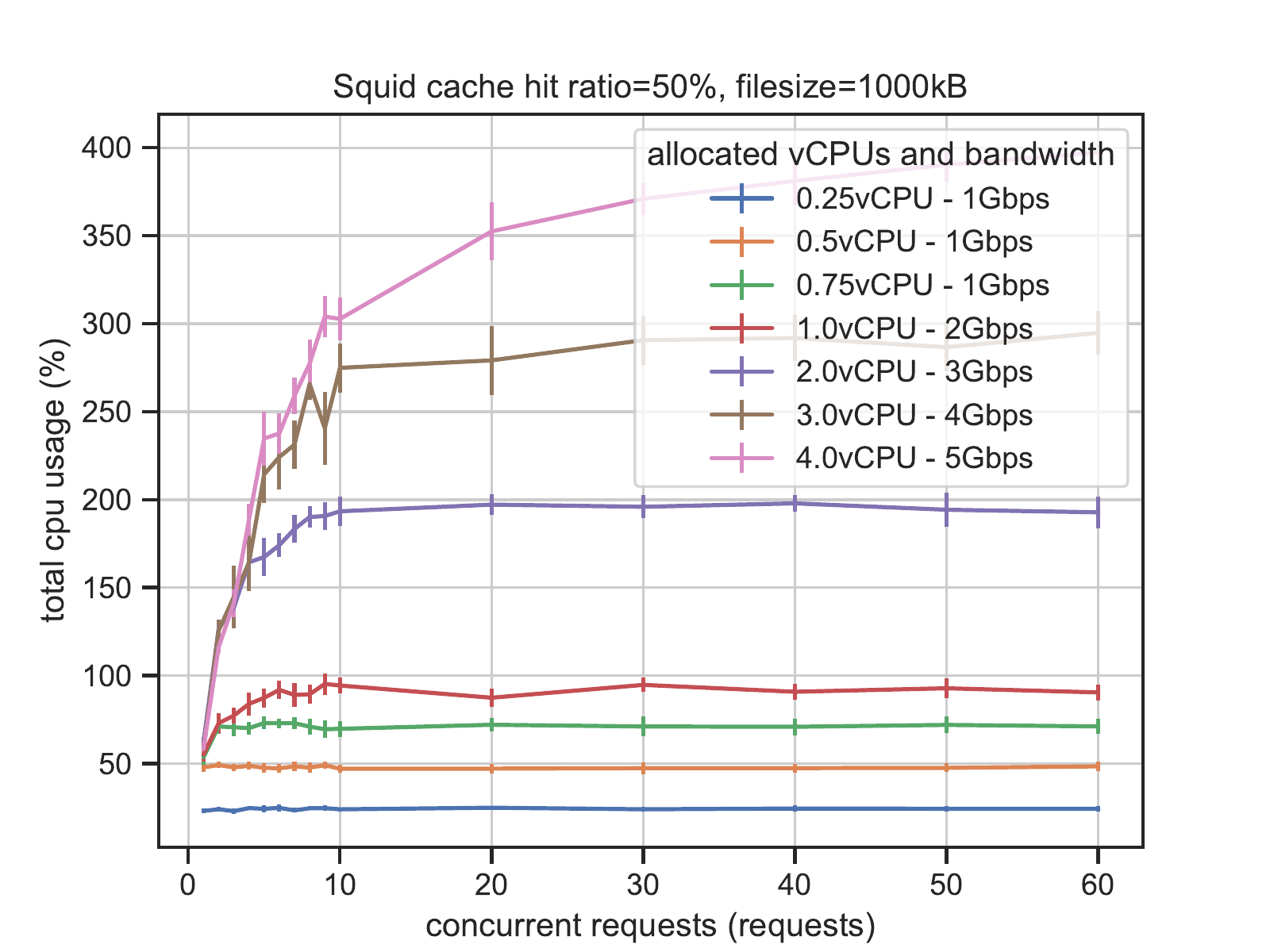}%
\label{cachecpu}}
\hfil
\subfloat[OVS packet loss]{\includegraphics[width=45mm]{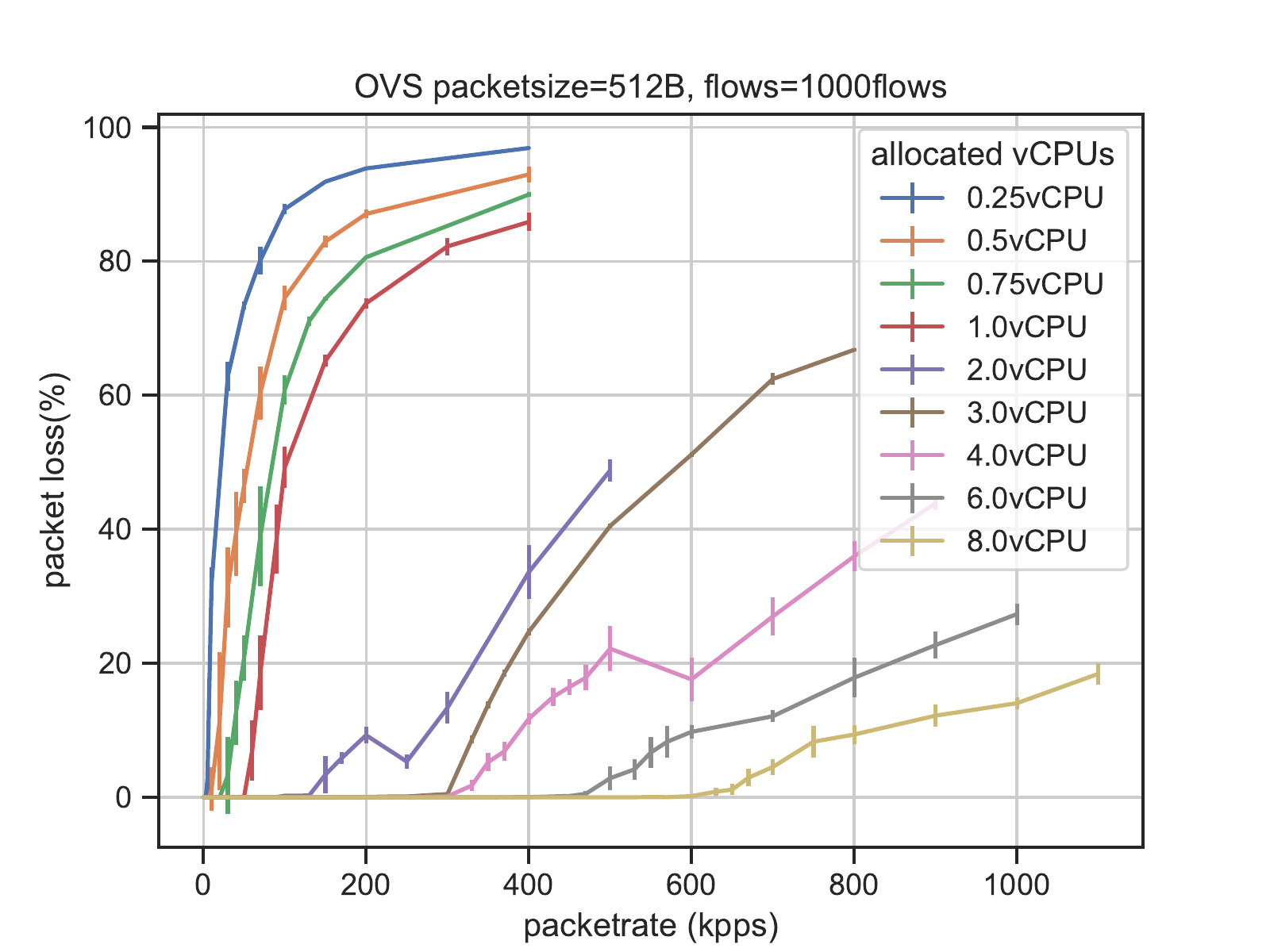}%
\label{ovsloss}}
\hfil
\subfloat[Router packet loss]{\includegraphics[width=45mm]{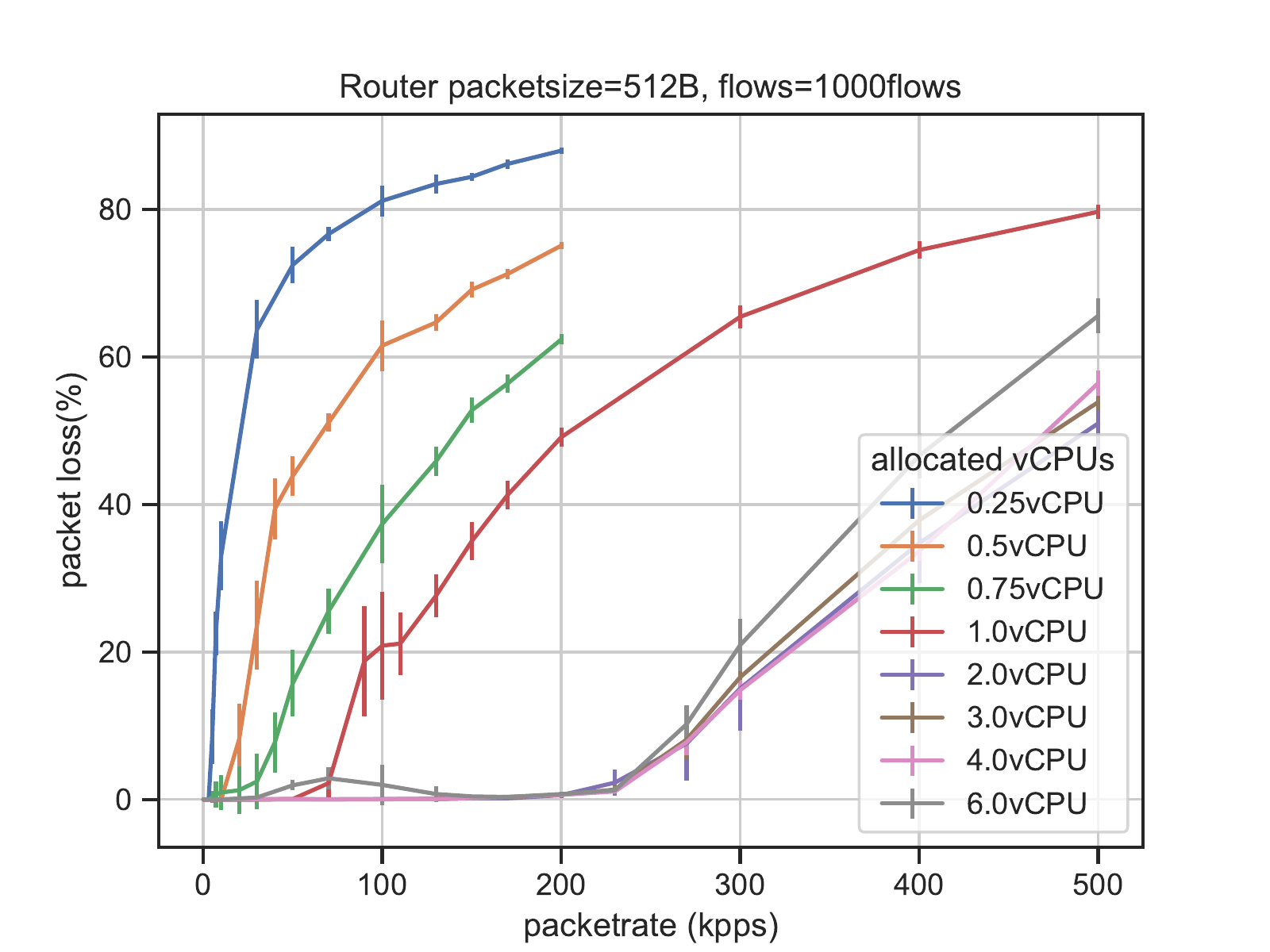}%
\label{routerloss}}
\hfil
\subfloat[Firewall packet loss]{\includegraphics[width=45mm]{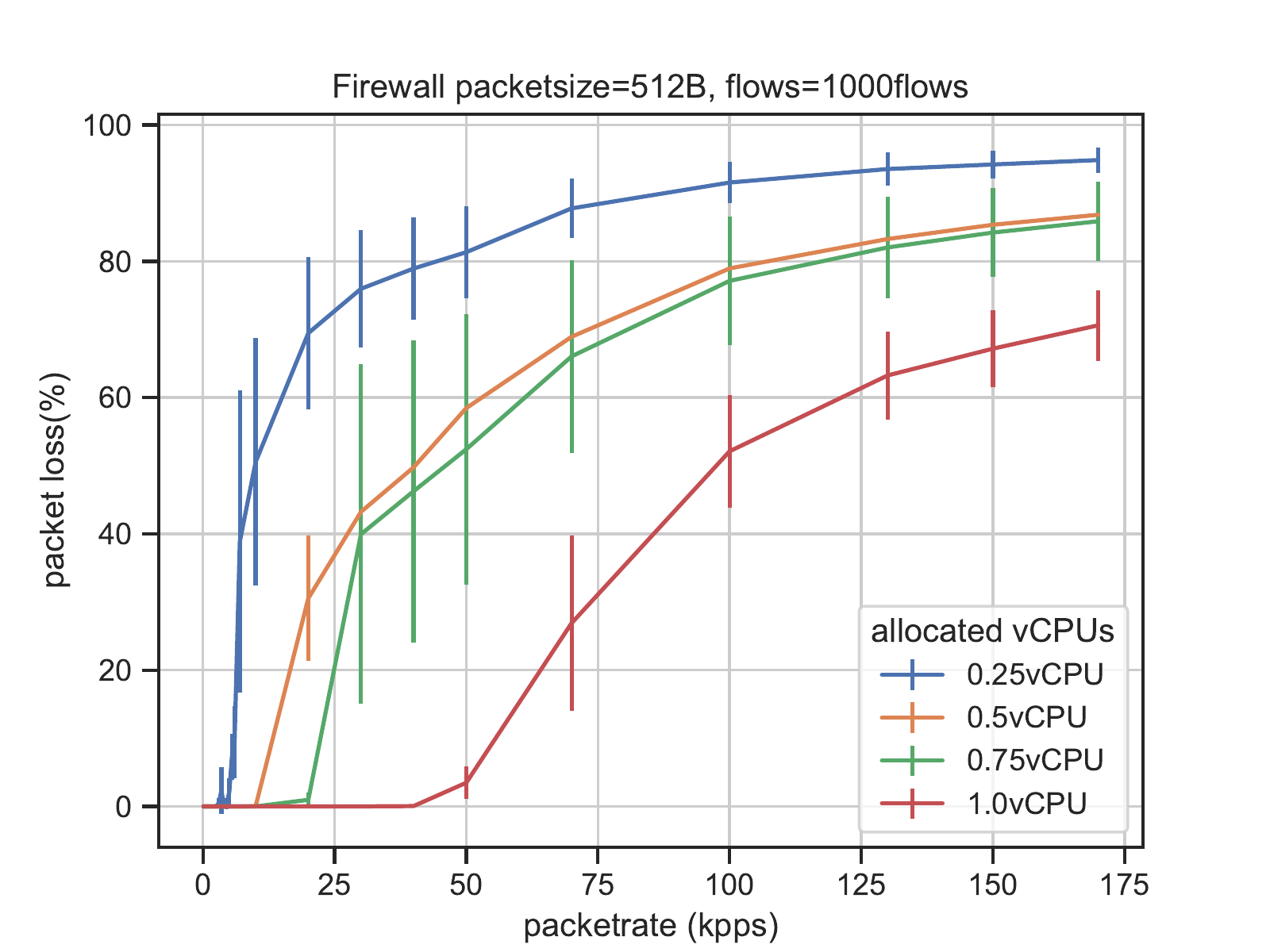}%
\label{fwloss}}
\hfil
\subfloat[Cache response time]{\includegraphics[width=45mm]{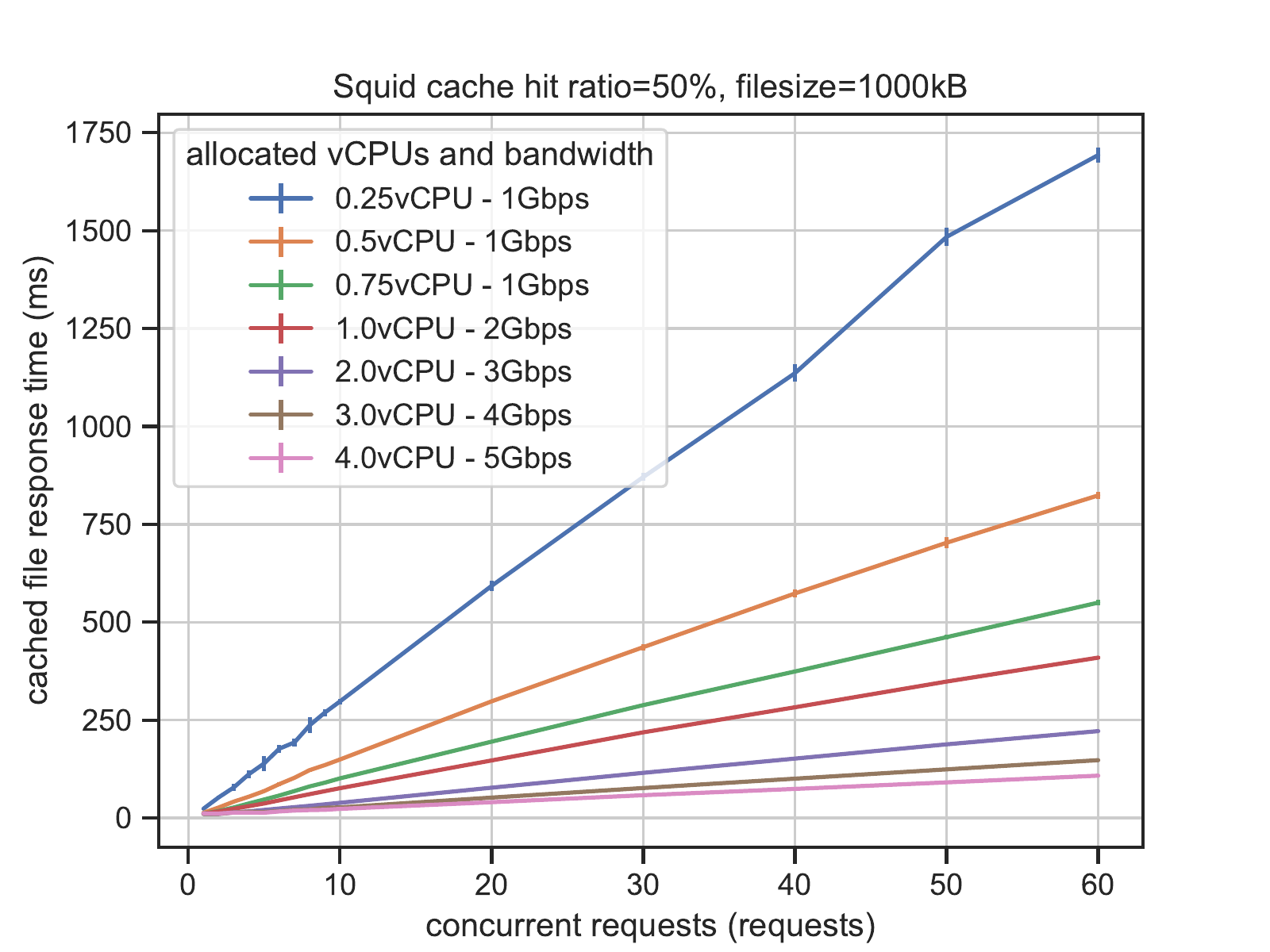}%
\label{cachedtime}}
\caption{\color{black}Subset of measured VNF metrics under different resource allocations (with 99\% confidence interval).}
\label{measurements1}
\end{figure*}

\subsection{Measured Network Functions}
\label{measurement_parameters}
Typical NFV use cases such as a virtualized Evolved Packet Core (EPC) or IP Multimedia Subsystem (IMS) make primarily use of request-based or packet forwarding VNFs \cite{ETSIusecases}.
Since we want to analyze the typical relation between workload, resource and performance metrics, we select four typical packet forwarding or request-based VNFs.
These VNFs exemplify our generic understanding of a VNF implementation:
the available amount of CPU and bandwidth is scheduled over the ingress workload, proportionally to the incoming packets or requests. So each processed packet or request will receive a fair share of the available resources. After resource saturation, performance starts to deteriorate, since packets and requests will receive a smaller share of resources as prior to saturation.
We suspect that the trends observed in these VNFs can fit to other ones also, as long as the VNF implementation is based on the same principle: resource usage is proportional to packet or request rate and KPIs are affected by resource saturation.

We monitored the performance of three typical packet forwarding-based VNFs:
\begin{itemize}[leftmargin=*]
\item \textbf{Router}: an evaluation license of a commercially General Available vRouter, implemented as a VM. In Fig. \ref{routerloss} we can see that the performance of this router is capped from 2vCPUs onwards. {\color{black} The traffic source and sink are each in a different L3 subnet. We do not alter the default routing table and let the router forward the traffic from source to sink.}
\item \textbf{Firewall}: an evaluation license of a commercially General Available vFirewall. 
{\color{black}This is a stateful firewall, providing L3, L4 and L7 functionality. We activate a built-in set of rules to protect against: (i) DoS, operating on L3 and L4. (ii) Additionally we  block SSH, SYSLOG and MYSQL services and (iii) we enable the offered DNS, Web and AntiVirus inspection tools operating at L7.} 
Implemented as a VM, the license limits the resource allocation to max 1vCPU. This is seen in Fig. \ref{fwcpu} and \ref{fwloss} where resource allocation stops at max 1vCPU.
\item \textbf{OVS}: To compare with the performance of the other packet forwarding VNFs, we deploy OpenvSwitch v2.10.1 (an open source softwarized vswitch implementation) into a VM based on Alpine Linux v3.9.1. The OpenvSwitch is configured as a standalone switch, so a flow entry is inserted for every unique flow passing through, one per unique mac source-destination pair. {\color{black} Note that to benefit from multi-core cpu allocation, we must enable multiqueue virtio-net drivers in KVM. This enables packet sending/receiving processing to scale with the number of available vCPUs of the guest VM. We must divide the available vCPUs over the specified number of queues in the virtio driver and the processing in the VM (OVS) itself.}
\end{itemize}
\medskip
We stress the above VNFs by generating multiple unique parallel flows. Also the packetsize is varied. The tool \textit{Scapy} is used to craft a .pcap file which describes a stream of packets with varying mac addresses.  \textit{Tcpreplay} is then used to stream the .pcap file at a given packetrate from the traffic source.
There is also an iperf stream running, with an iperf server in the traffic sink. This is used to monitor packet loss.
For the router {\color{black} and firewall} to function properly, we need to make sure the ARP table of the VNF contains the mac addresses of the generated packets, so the router forwards the packets properly to the traffic sink. This is done by arp spoofing the VNF under test from the traffic sink.

\textit{Generated workload metrics}: 
\begin{itemize}
\item packetrate: [0.1-1000]kpps, 20 different packetrate values are selectively chosen, spaced evenly along the log scale.
\item packetsize: [64,128,512,1024,1500]bytes 
\item unique flows: [1,2,10,100,1000,10000] parallel flows
\end{itemize}

\textit{Resource metrics}: 
\begin{itemize}
\item CPU allocation: [0.25, 0.5, 0.75, 1, 2, 3, 4, 6, 8] vCPUs
\end{itemize}
A subset of the measured configurations is illustrated in Fig. \ref{ovscpu} \ref{routerecpu}, \ref{fwcpu}. Some VNFs have a more limited set of vCPU allocations, due to license issues.
Since we specify the generated packetrate and packetsize up front, the bandwidth requirements for the VNF can be easily determined.

\textit{Performance metric}: \\
We choose packet loss (\%) as the main KPI to reflect the performance of the packet forwarding VNFs. 
A subset of the measured configurations is illustrated in Fig. \ref{ovsloss} \ref{routerloss}, \ref{fwloss}.

\medskip
We also check the performance of a typical request-based VNF. For this type of VNFs, both the allocated bandwidth as the number of vCPUs influence the performance.
\begin{itemize}[leftmargin=*]
\item \textbf{Cache}: Squid v3.5 is deployed in a Docker container. Following the recommended configuration options for multi-threaded performance we deploy one Squid instance per allocated vCPU core, each instance serving at a different TCP port. The incoming requests are then load-balanced over the Squid instances using rules in \textit{iptables}. We configure Squid as a cache server, using RAM to store the cached files.
\end{itemize}
\medskip
The cache server is serving files to the traffic generator. 
We stress the cache server by generating $n$ concurrent file requests. This means at any given time, there are $n$ pending file requests ongoing, by $n$ threads.
\textit{Locust.io} is the tool used to generate the file requests. 
The traffic source is generating file requests of varying filesizes.
The traffic sink is a webserver, a Python based implementation which generates a random file with the requested size.
Files which are not cached in the DUT are requested to the traffic sink webserver. We also control how much of the file requests ask non-cached files by manipulating the http request header (setting `Cache-Control: no-cache` in the http header). 
This emulates a varying cache hit ratio.

\textit{Generated workload metrics}: 
\begin{itemize}
\item concurrent requests: [1-60], 15 different values are uniformly chosen.
\item filesizes: [1,5,10,50,100,500,1000]kB 
\item cache hit ratio: [10,50,90]\%
\end{itemize}

\textit{Resource metrics}: 
\begin{itemize}
\item CPU allocation: [0.25, 0.5, 0.75, 1, 2, 3, 4] vCPUs
\item Bandwidth allocation: [0.25-5]Gbps
\end{itemize}
To model the influence of the bandwidth limit, a number of bandwidth allocations is dynamically chosen, relative to the vCPUs used. For each allocated number of vCPU, we allocate up to three selected bandwidth limits (e.g. for 0.25vCPU we test 0.25 and 1Gbps, for 2vCPU we test 1, 2 and 3Gbps). The bandwidth allocation is configured by a ratelimit on the download link to the traffic source, using \textit{tcset}.
A subset of the measured configurations is illustrated in Fig. \ref{cachecpu}.

\textit{Performance metric}: \\
We choose the response time of cached file requests (ms) as the main KPI to reflect the performance of the cache server, 
as illustrated in Fig. \ref{cachedtime}.

\medskip
For all the tested VNFs we consider CPU as the most important resource metric, and assume CPU is more likely to become a bottleneck resource than memory. 
This is also confirmed in \cite{kulkarni2017nfvnice}.

\medskip
Taking all the different workload and resource configurations into account, there are up to 5400 different configurations to be tested for one VNF. We reported in section \ref{stability} that each configuration takes an average of 30s to get a stable measurement, this means the complete profiling can take up to 45hours for one VNF. The iteration through all the profiled configurations is automated by the Profiling Controller, which controls the above mentioned tools and settings to generate all the different workloads and allocate the specified resources. This long profiling time pinpoints one of the main problems with VNF profiling: many possible configurations lead to a multiplicative growth rate of the test time. Possible ways to mitigate this include narrowing down the possible configuration options, or apply a better way to select which sample configurations are most interesting to measure.

\begin{figure*}[!t]
\centering
\subfloat[Observed generic trends to model]{\includegraphics[clip=true, trim=5 350 700 5, width=.33\textwidth]{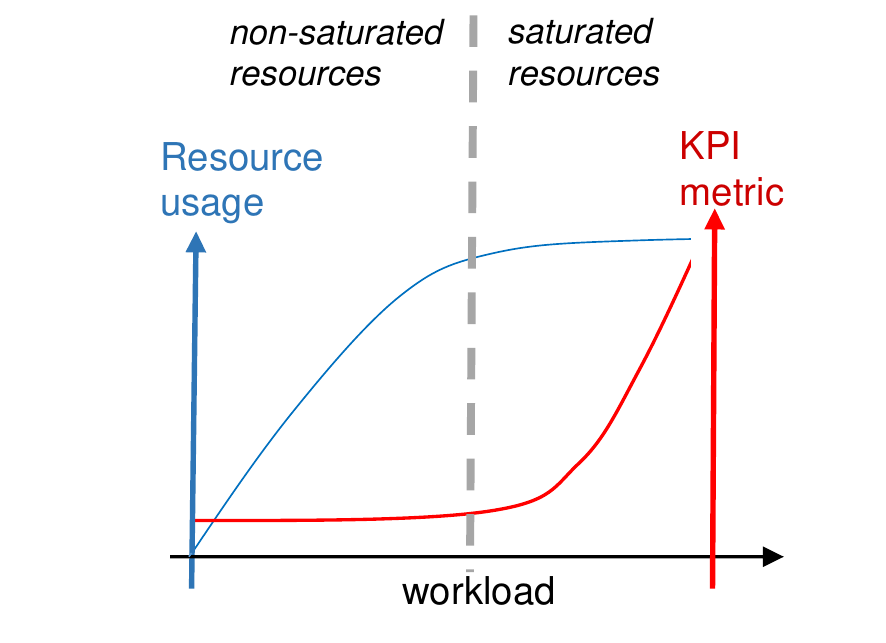}%
\label{abstract_trends}}
\hfil
\subfloat[OVS example subset]{\includegraphics[width=.33\textwidth]{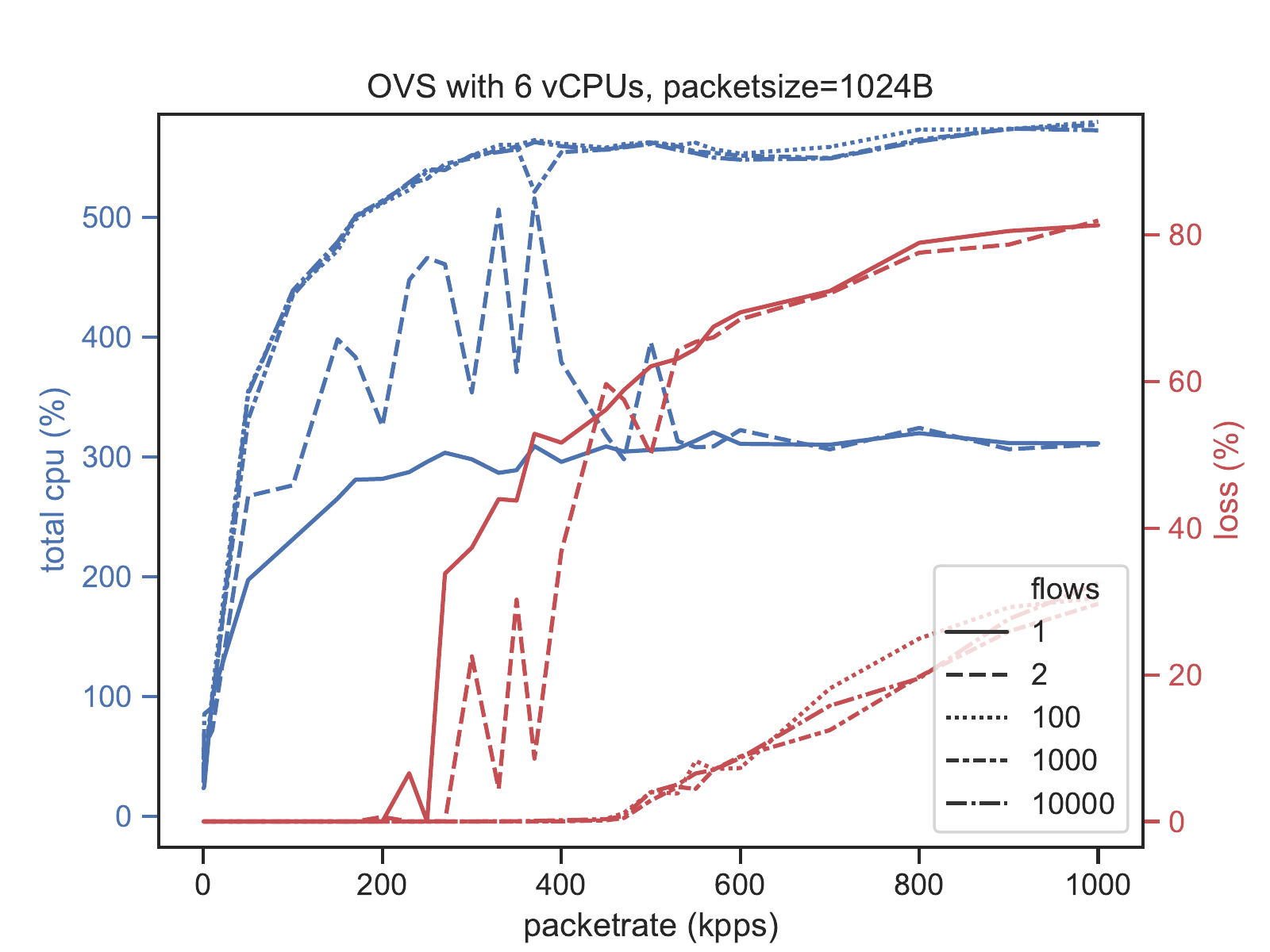}%
\label{variation_eg1}}
\hfil
\subfloat[Cache example subset]{\includegraphics[width=.33\textwidth]{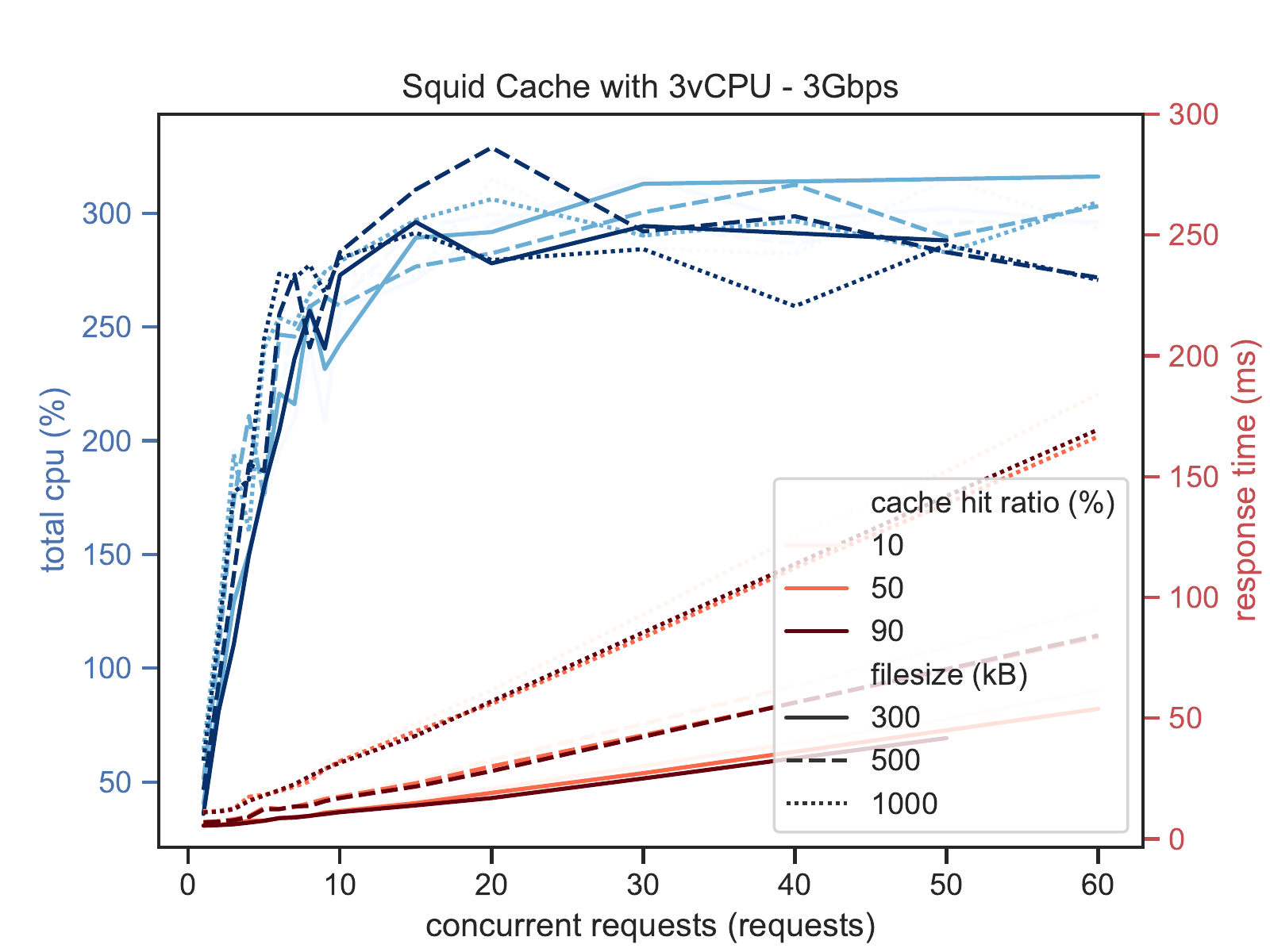}%
\label{variation_eg2}}
\caption{\color{black}Generic observed trends and data example subsets. This shows how varying workloads induce noisy, non-linear resource and KPI measurements which follow certain trends.}
\label{variation_compare}
\end{figure*}

\section{Analysis of Profiled Data}
\label{analysis}
In the previous section we described how different VNF metrics were gathered under a set of different workload and resource configurations.
Having this profiled information available, we want now to know, how well we can predict the performance of the VNF from this dataset by a trained model. We try out several modelling techniques and compare their accuracy.

In Fig. \ref{measurements1} we can see a subset of the resource allocations and the related performance measured at the same time. Every data point is the mean of minimum {\color{black}15} repeated measurements. We also derive a confidence interval for each point.
Although some noise is present, we can clearly identify some trends in the profiled datasets:
\begin{itemize}[leftmargin=*]
\item When resources are freely available (no CPU starvation), workload and resource usage are highly correlated. CPU usage rises with increasing workload on the x axis. Meanwhile, performance metrics on the bottom row plots remain fairly constant while CPU still has margin.

\item When resources become scarce (CPU reaches saturation), CPU usage flattens to the maximum available amount, even if workload still increases. The correlation between measured workload on the x axis and resource usage is lowering. From this point onwards, the performance metrics start to increase more rapidly. Now, workload and performance metrics are more correlated.
\end{itemize}

It is important to notice here that the same kind of trend is witnessed with every VNF. The steepness and trend break-point of the monitored resource usage and performance differs from configuration to configuration.
{\color{black}
This is illustrated in Fig. \ref{variation_eg1} and \ref{variation_eg2}.
For better readability, we only show a small subset of processed workloads and the  according CPU usage and KPIs.
During our measurements, we observe a general trend as depicted in Fig. \ref{abstract_trends}
(this trend can also be seen in Fig. \ref{measurements1}).
Intuitively, this also corresponds to how we expect a VNF to be generally implemented: with resource saturation reached by increasing packet or request rate.  
While the data in Fig 3 is averaged over 15 repeated tests, the remainder of the paper is based on a limited subset of only five repeated tests. We do this to reflect more a real-life situation where limited time is available to gather many repeated tests. As a result, a certain portion of noise is not averaged out in the data as seen in Fig. 4. 
In general, the gathered data is characterized by following aspects, which impact the accuracy of the later used modelling methods:
}

\paragraph{Heavy non-linear relations and trends in various monitored metrics}
There is a steep trend-break in the performance when the resources reach saturation. Also the variation caused by changing workloads cannot be fit to low-grade polynomials as is mostly highly non-linear. This causes regression methods to fail at modelling the performance trends of the VNFs as regression methods try to fit the performance to a polynomial combination of workload and resource metrics. This is shown in Fig. \ref{variation_eg1} and \ref{variation_eg2} where we see that the performance trends (red curves) are varying under different workload configurations. This variation seems hard to capture accurately in a model.

\paragraph{Noisy measurements of the performance metrics}
Due to noise on performance measurements, there is no guarantee for a monotonous trend for the measured performance. The noise can cause highly over-fitted models, especially with interpolation methods. (see e.g. Fig. \ref{variation_eg1} where a non-smooth curve would result from interpolating between measured performance values.)

\paragraph{Inability to gather a lot of training samples because of the slow profiling measurements}
Due to time restrictions, the amount of profiled samples is rather limited. We need to work with a limited set of samples where only selected workload and resource configuration have been measured before. This can give problems with interpolation and nearest neighbor based methods, if little neighboring samples are available. Also machine learning based methods such as ANN tend to fail with little training samples.

{\color{black}
\paragraph{A monotonic function approximates the observed trends}
When averaging multiple repeated measurements, a smoother curve occurs as illustrated in Fig. \ref{measurements1} and \ref{abstract_trends}. This is useful information because we can use the smoother curve to approximate the noisy data. 
Moreover, we can use a monotonic function which helps to calculate a resource recommendation as outlined further in Section \ref{recommendation}. In our algorithm, the observed monotonic function simplifies the process to lookup the according workload for a specified resource and KPI value, since monotonicity avoids the need to take local extrema into account. 
}

\begin{figure*}[!b]
\centering
\subfloat[Regression]{\includegraphics[width=.33\textwidth]{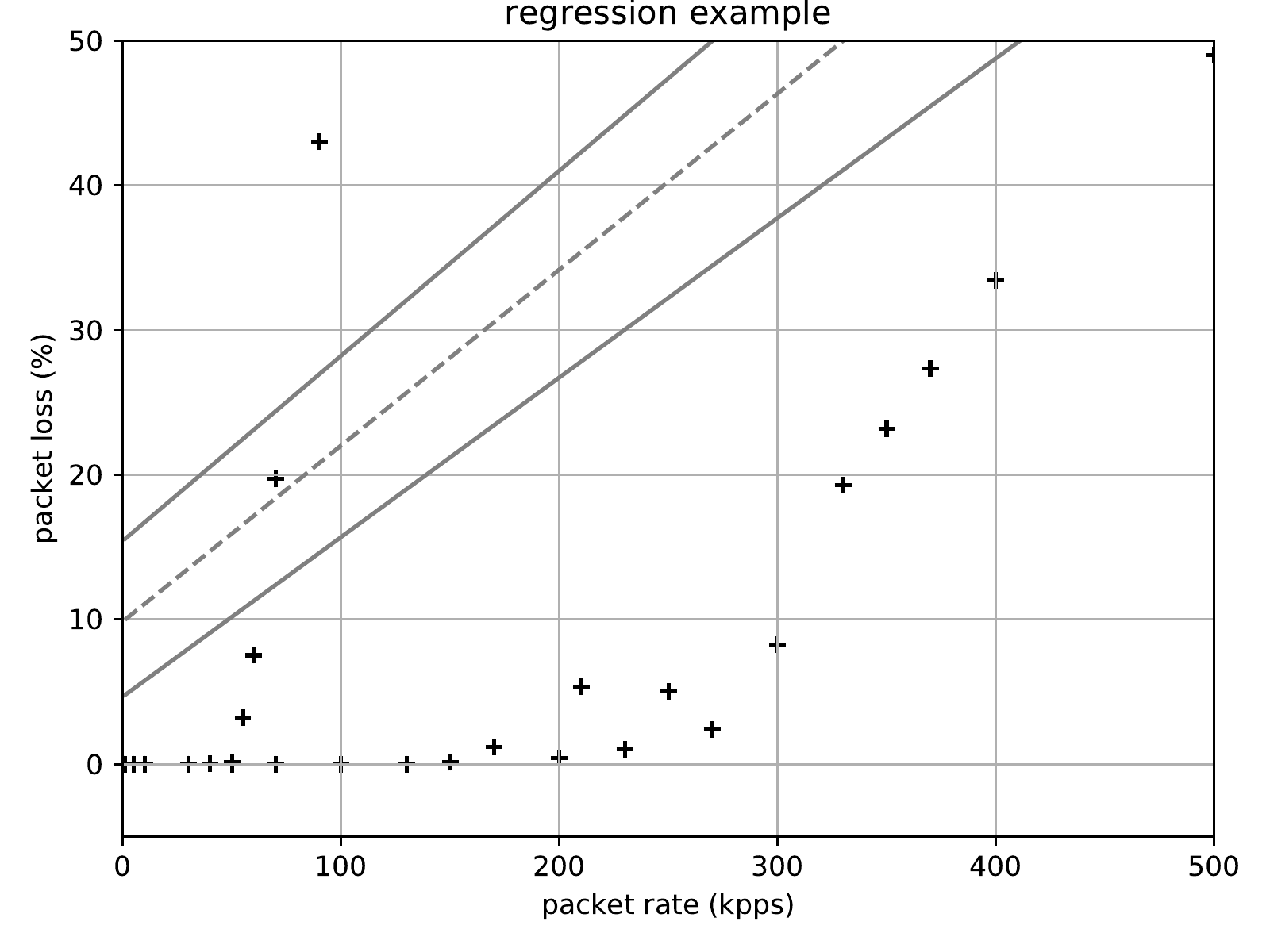}%
\label{regr_eg}}
\hfil
\subfloat[kNN]{\includegraphics[width=.33\textwidth]{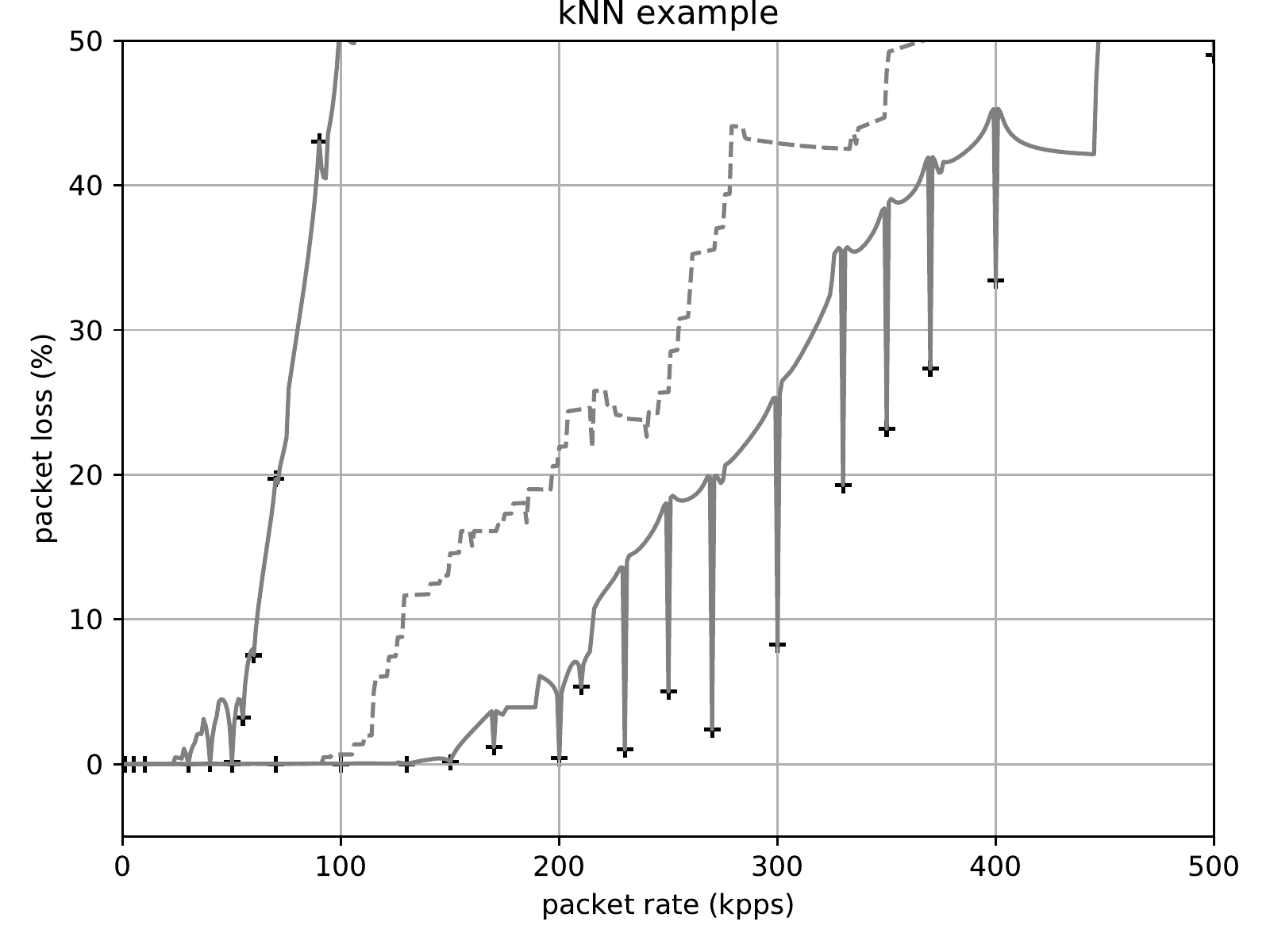}%
\label{kNN_eg}}
\hfil
\subfloat[Interpolation]{\includegraphics[width=.33\textwidth]{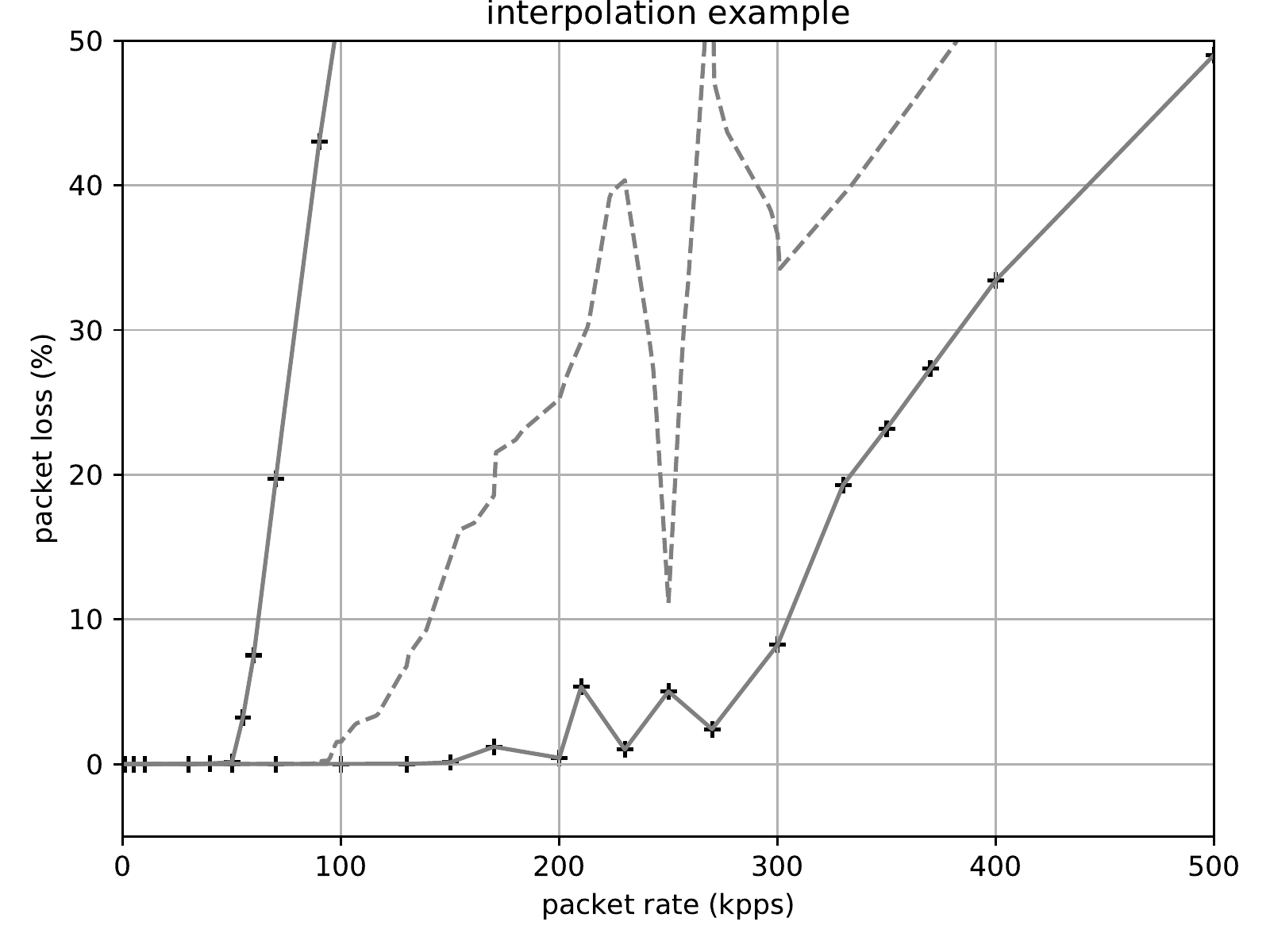}%
\label{interp_eg}}
\hfil
\subfloat[ANN]{\includegraphics[width=.33\textwidth]{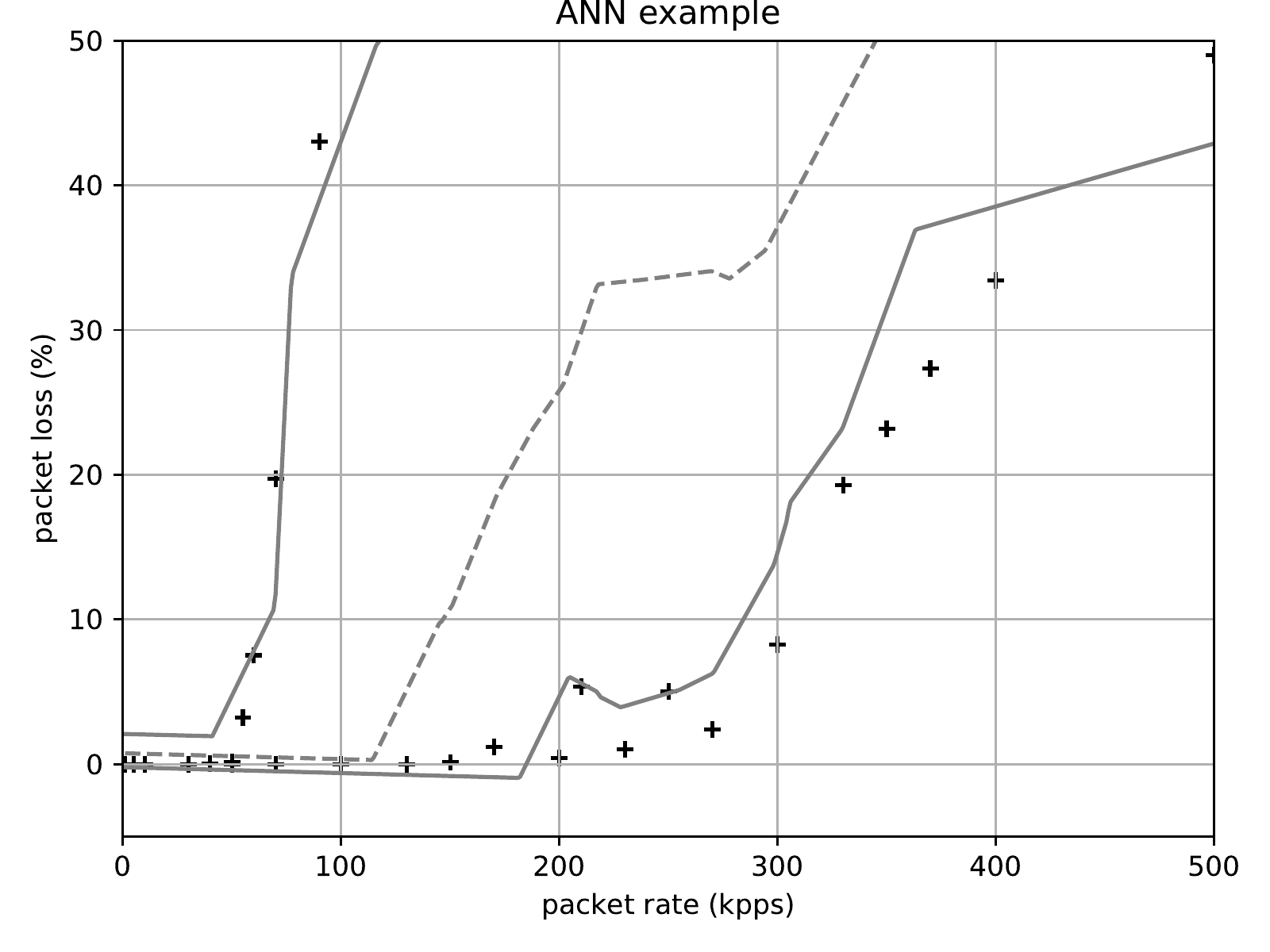}%
\label{ANN_eg}}
\hfil
\subfloat[Curve Fitting]{\includegraphics[width=.33\textwidth]{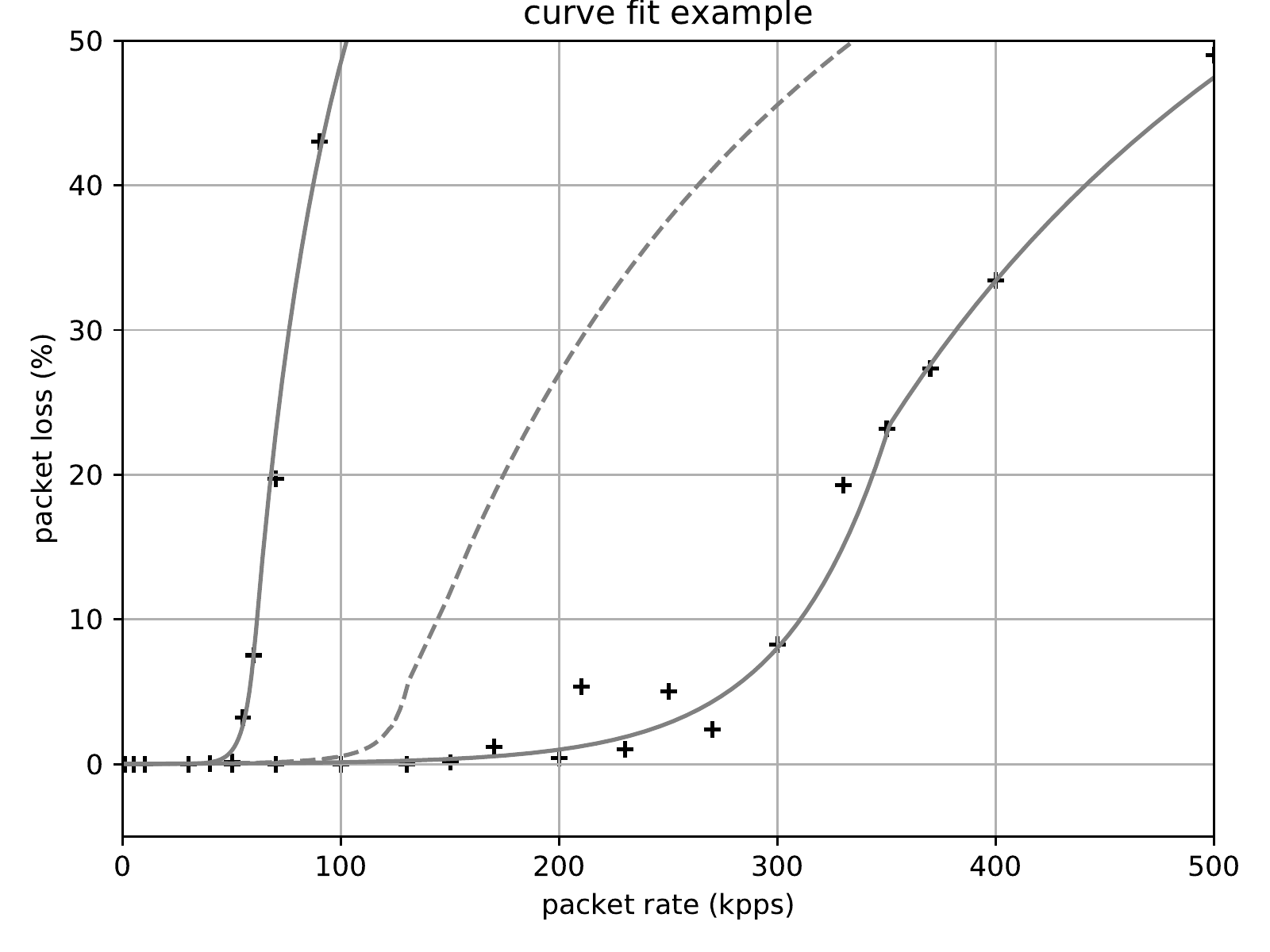}%
\label{curve_fit_eg}}
\hfil
\subfloat[]{\includegraphics[clip=true, trim=0 220 480 0, width=.33\textwidth]{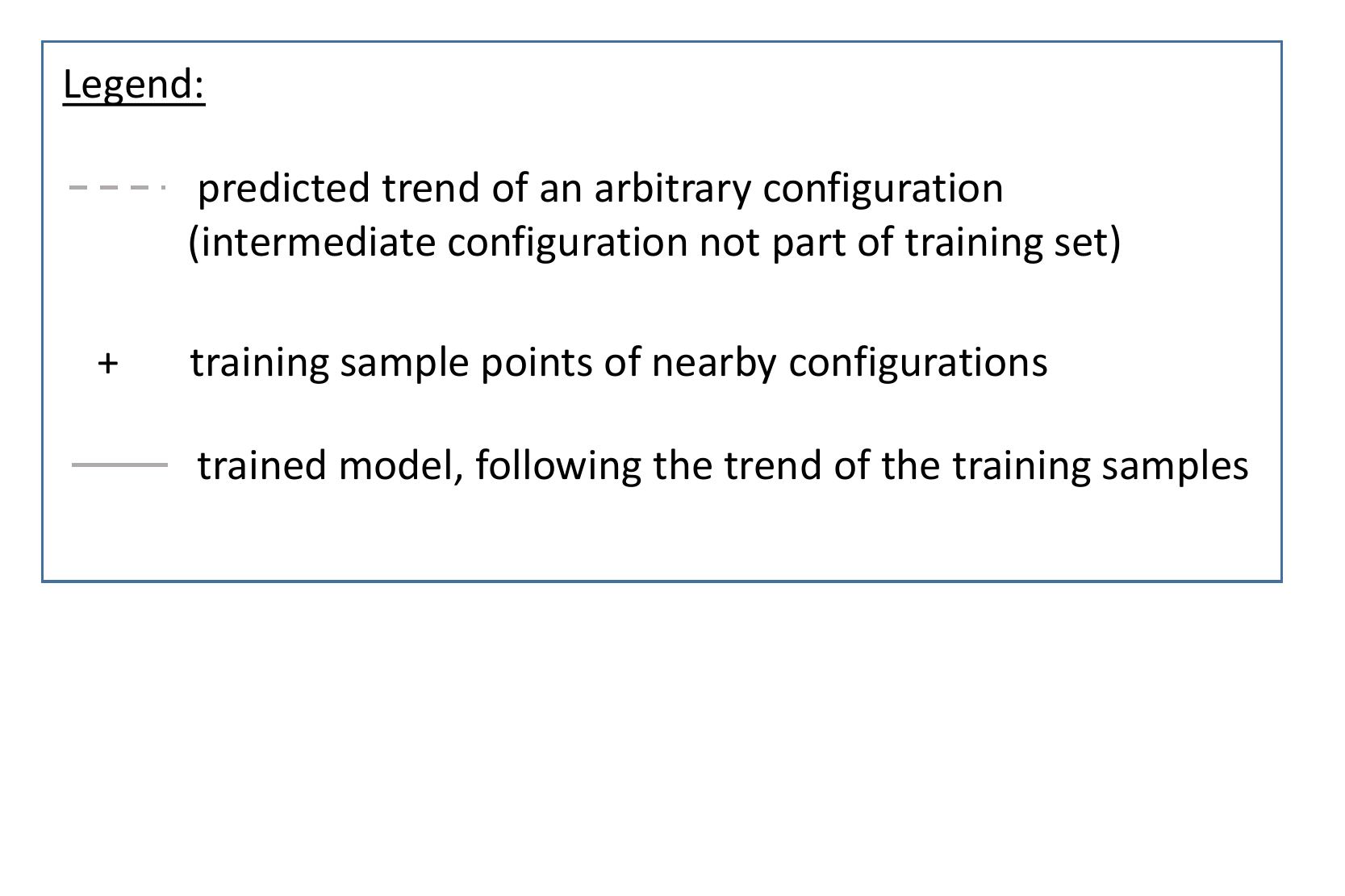}%
\label{fig5_legend}}
\caption{\color{black}Comparison of different modelling techniques fitted to the same sample set. Each plot depicts three lines, i.e. the same three configuration settings. The dashed line shows how the trained model predicts a configuration outside of the training set. Curve fitting approximates best the observed trends.}
\label{model_compare}
\end{figure*}

\subsection{General Analysis Method}
\label{general_analysis}
For a user of a network service, the performance of the deployed VNFs is specified in the SLA. Typically the SLA defines performance limits which must be met while the workload varies in a certain range. For example, the VNF can have max 1\% packet loss while the incoming traffic is max 1Gbps, or the response time is max 500ms while the incoming request rate is 10 requests/sec. So the SLA imposes a relation between incoming workload and performance KPIs. The operational platform however, can only allocate resources to a VNF, it has no direct idea how the allocated resources impact the performance of the VNF. This is where the VNF profile can help. The profiled dataset can be used to derive a relation between allocated resources and resulting performance. The main goal of doing the profiling measurements, is to derive a model which predicts the needed resource allocation, in function of the specified workload and performance in the SLA.
From an abstract and generalized viewpoint, the VNF performance model can be described as:

\begin{equation} \label{fmodel1}
f(wl, res) = perf
\end{equation}
where:\\
$wl$  = input workload  (e.g. packetrate, filesize) \\
$res$ = resource allocation (e.g.number of allocated vCPUs)\\\
$perf$ = VNF KPI metrics (e.g. packet loss)\\

This model $f$ allows us to predict the performance at any given workload and resource allocation. Next, we need to find a resource configuration which meets our performance target:  $perf_{target}$. Therefore we can define following cost function:

\begin{equation} \label{fmodel2}
minimize  \left| f(wl, res) - perf_{target} \right|
\end{equation}

The objective is now to find the minimal (cheapest) resource allocation which can process a given workload at a given performance target. We do this by iterating over all profiled resource allocations, and in each resource allocation we find the maximum workload which minimizes the above cost function. 
As a result, we derive from the profiled dataset how much workload the VNF can process under given resource allocations. This can be used by the orchestration or scaling procedure to estimate the optimal resource allocation in order to process a certain specified workload with known performance. We will exemplify this procedure in section \ref{recommendation}. {\color{black} As mentioned earlier, a smooth monotonic function $f(wl, res)$ would simplify the calculation, since no multiple local minima have to be taken into account.}

\subsection{Model Comparison}
\label{model_comparison}

The function $f$ in Eq. \ref{fmodel1} and \ref{fmodel2} can be implemented using various techniques. In this section we describe our learnings from comparing following methods:
\begin{itemize}[leftmargin=*]

\item \textbf{Linear Regression}: We can try linear regression methods to fit to non-linear trends by using polynomial expansion on the predictors of the model. In this case this means including also the mutual products of workload and resource allocation parameters and even include higher order products. As exemplified in \cite{huang2010predicting} the introduced collinearity is then handled by using the Lasso method to select only the most relevant terms in the regression. The result is however not satisfactory. And we conclude that regression works not well in this use-case. 

\item \textbf{k-Nearest Neighbors (kNN)}: By taking the average of the k nearest profiled samples, we can model non-linear trends more easily. We search the optimal k, by testing different values for k and checking which one yields the best accuracy. We also standardize the configuration metrics, so the distance to neighboring configurations is not skewed by the different scales of the metrics.

\item \textbf{Interpolation}: Instead of calculating the mean of neighboring samples, we can also interpolate between surrounding samples. The interpolant is constructed by triangulating the input data using Delaunay triangulation, and on each triangle performing linear barycentric interpolation. This method also works in multiple dimensions, so we can interpolate between any number of configuration metrics to predict the performance of an intermediate configuration.
We use the method \textit{griddata} implemented in the Python SciPy library\cite{scipy}.

\item \textbf{Artificial Neural Network (ANN)}: Neural networks are widely applied to model non-linear datasets and we train an ANN to model the shown performance trends. 
{\color{black}The used ANN type is a multi-layer perceptron regressor (using the standard relu activation function). 
The hyperparameters are found by exhaustively testing different values for optimal accuracy (each VNF yields different model parameters). The regularization parameter $\alpha$ ranges from $10^{-1}$ to  $10^{-6}$.
We obtain the best results when using two hidden layers, within each hidden layer a number of nodes varying between 15 and 20. Higher numbers of hidden layers and nodes give no further improvement. For each VNF, the input layer has a node for each workload and resource configuration parameter, the output layer has one node for the used KPI metric. 
}

\item \textbf{Curve Fit}: We try to fit a pre-defined set of analytic curves to the measured performance samples for each profiled configuration. {\color{black} The performance values of} any new configuration are then interpolated between the fitted curves (using the same interpolation method as described above). This method offers the best accuracy, as we will further detail in section \ref{curve_fit_method}.

\end{itemize}

\vspace{1mm}
\subsubsection{General Thoughts on the Models Used}
{\color{black}
In Fig. \ref{model_compare}, each plot shows the same subset of measured samples and how they are approximated. 
Here we can compare how well each method succeeds at modelling the smooth monotonic function we put forward as objective.
The 'trained configurations' are defined in Section \ref{measurement_parameters} and comprehend the total set of profiled resource allocation and workload settings.
The dashed line shows how the model predicts the performance of an 'untrained configuration'. An 'untrained configuration' means this resource/workload combination is not tested in the profiled dataset, hence no samples are available. The model must learn the behaviour of any untrained configuration from the limited set of profiled training configurations.

The regression (Fig. \ref{regr_eg}) is the least accurate becuase it cannot handle the steep trend break happening at resource saturation.
We can also clearly see how the ANN, kNN and Interpolation method do not guarantee a monotonic rising function. 
The Curve Fit method approximates best the trends seen in previous plots (Fig. \ref{measurements1} and \ref{variation_compare}), especially considering the imperfections of the data (noisiness and limited quantity of samples). 
Moreover, we can guarantee the modelled performance trend to be monotonous. This benefits  Eq. \ref{fmodel2}, as it would guarantee a single possible solution for the recommended resource allocation.
}

Table \ref{tab:accuracy} summarizes different accuracy metrics for the different methods per VNF. Each reported accuracy is the result of a 5-fold cross validation: The profiled dataset was divided into five equal parts, with each part serving once as the test set and the other parts forming the training set of the model.
In the last row we have normalized and averaged the accuracy metrics to be able to compare between the different VNFs. The Curve Fit method seems to be the overall winner with the lowest error values.

While the accuracy metrics for the ANN in Table \ref{tab:accuracy} might seem acceptable, care must be taken: as seen in Fig. \ref{ANN_eg}, there is no guarantee that the ANN models the samples in a sensible way. This means: packet loss should be zero at low packetrates and then monotonically increase. 
The same is true for the kNN method. For the Interpolation method (shown in \ref{interp_eg}), the monotonicity is broken by noise in the sample measurements.
The method we look for, should be able to yield a 'smoother' curve, which can model a fairly constant performance value at low workloads and then transition into a steeper curve.
To tackle the issues which decrease the accuracy in the above described methods, we develop a model based on curve fitting. As can be seen in Fig. \ref{curve_fit_eg}, this method guarantees a smooth and monotonically rising modelled performance trend.
We will detail the accuracy metrics later in section \ref{accuracy}. First we explain the Curve Fit method more in detail.

\begin{table*}[!t]
  \centering
  \begin{tabular}{|@{}c|@{}c|@{}c|@{}c|@{}c||@{}c|@{}c|@{}c|@{}c||@{}c|@{}c|@{}c|@{}c||@{}c|@{}c|@{}c|@{}c||@{}c|@{}c|@{}c|@{}c|}
  	\hline
           & \multicolumn{4}{c||}{Regression} & \multicolumn{4}{c||}{kNN}  &  \multicolumn{4}{c||}{Interpolation} %
           &  \multicolumn{4}{c||}{ANN} &  \multicolumn{4}{c|}{Curve Fit} \\  \hline 
           VNF & r2   & MAE  & MAD  & RMSE        & r2 & MAE & MAD & RMSE      & r2 & MAE & MAD & RMSE %
           & r2 & MAE & MAD & RMSE & r2 & MAE & MAD & RMSE\\ \hline 
   OVS(\%)    & 0.45 & 19.49 & 17.54 & 23.4      & 0.79 & 7.42 & 1.38 & 14.7  & 0.87 & 4.67 & 0.52 & 11.4 %
           & 0.93 & 4.95 & 1.88 & 8.42 & 0.97 & 2.33 & 0.48 & 4.75\\
   Router(\%)  & 0.53 & 14.87 & 11.71 & 19.01     & 0.95 & 2.06 & 0.13 & 5.99  & 0.97 & 1.52 & 0.11 & 4.69 %
           & 0.98 & 1.51 & 0.24 & 4.1 & 0.97 & 1.95 & 0.28 & 4.92\\
   Firewall(\%)& 0.85 & 8.21 & 5.54   & 11.78     & 0.98 & 1.73 & 0    & 4.23  & 0.99 & 1.51 & 0    & 3.36 %
           & 0.99 & 1.18 & 0.12 & 2.63  & 0.99 & 1.65 & 0.07 & 3.72  \\   
   Cache(ms)   & 0.85 & 14.8 & 2.15 & 47.98       & 0.92 & 6.86 & 0.27 & 36.29 & 0.95 & 3.98 & 0.13    & 28.31 %
           & 0.99 & 5.16 & 1.25 & 12.49 & 0.98 & 1.84 & 0.18 & 19.96\\ \hline
Norm.Avg. & 0.67 & 1 & 1 & 1     & 0.91 & 0.3 & 0.05 & 0.51  & 0.95 & 0.2 &0.02  &0.4  &0.97  &0.21  &0.18  &0.26  &0.98  &0.14  &0.04  &0.3\\ \hline
  \end{tabular}
  \medskip 
  \caption{Accuracy metrics of the investigated modelling techniques.(MAE, MAD and RMSE are in ms for the Cache VNF, packet loss (\%) for the other VNFs.)}
  \label{tab:accuracy}
\vspace{-5mm}
\end{table*}

\begin{figure*}[!b]
\centering
\subfloat[Firewall curve fitting (Eq. \ref{packet_fwd})]{\includegraphics[width=.5\textwidth]{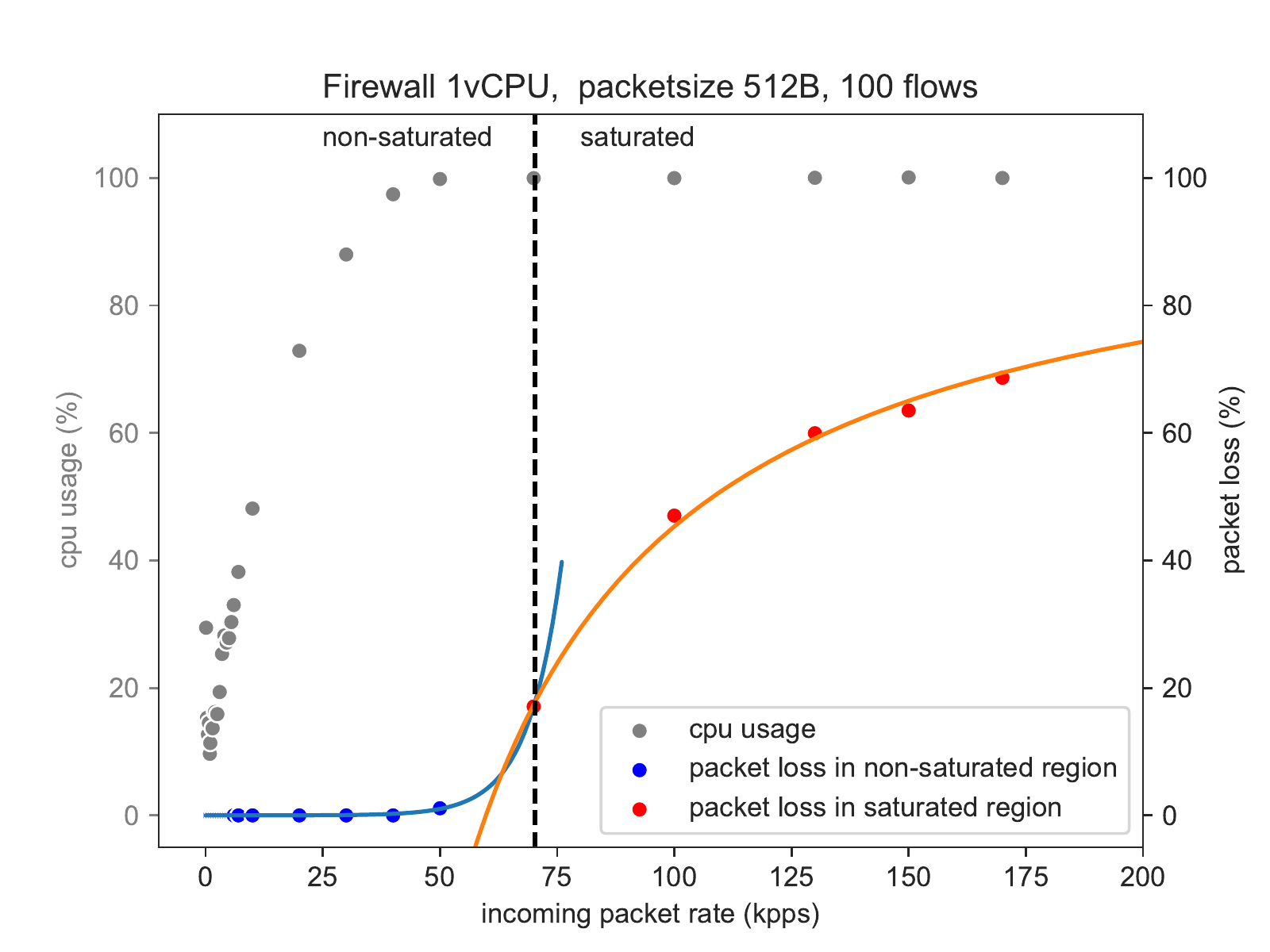}%
\label{curvefit_fw}}
\hfil
\subfloat[Cache curve fitting (Eq. \ref{vnf_reqs})]{\includegraphics[width=.5\textwidth]{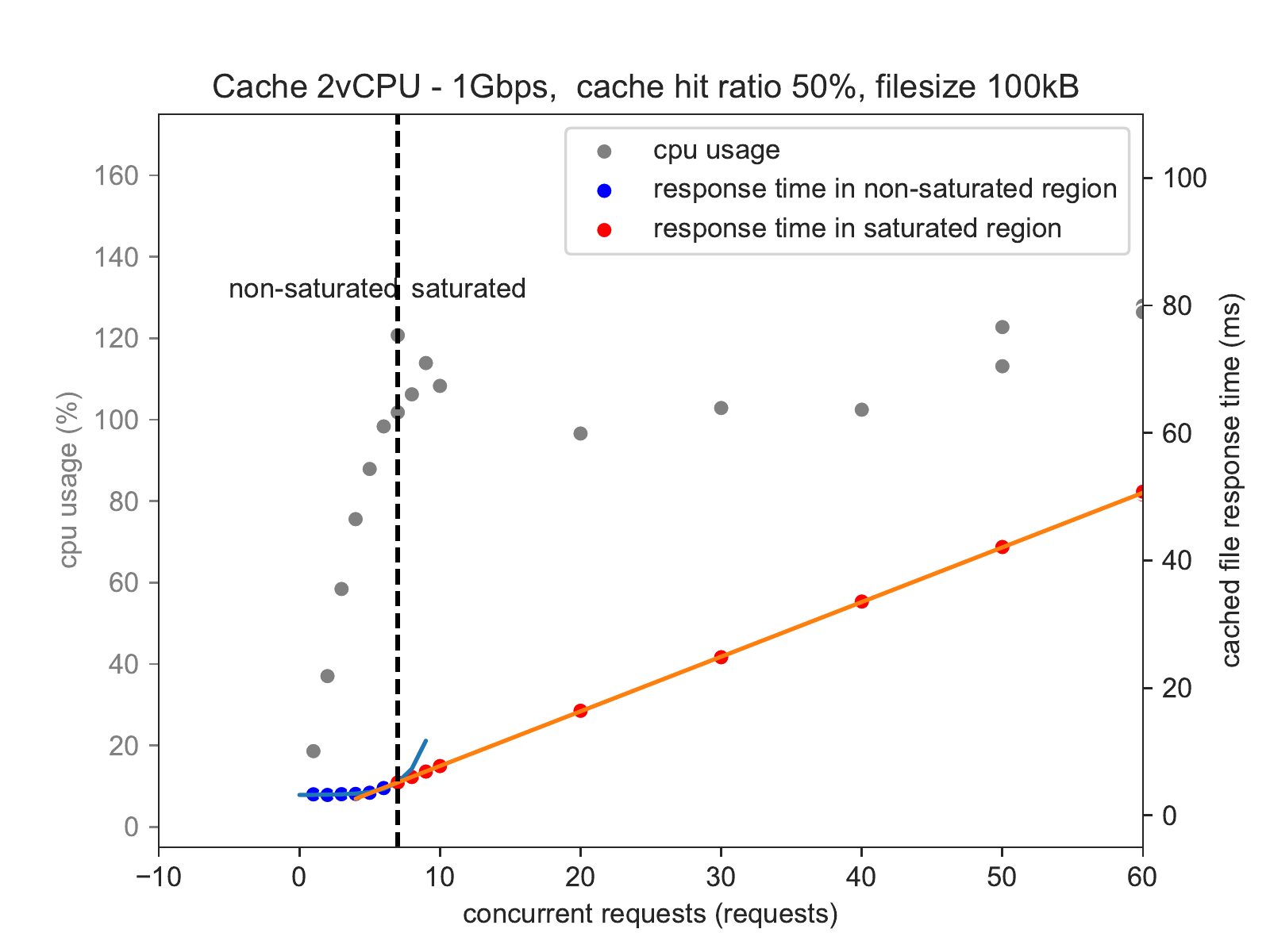}%
\label{curvefit_cache}}
\caption{\color{black}Curve fit strategy exemplified in two sample subsets. Different trend curves are used before and after resources get saturated.}
\label{curvefit_detail}
\end{figure*}

\subsection{The Curve Fit Method}
\label{curve_fit_method}
The training procedure of our Curve Fit method is illustrated with two sample subsets in Fig. \ref{curvefit_detail}.  
Based on observations of the profiled dataset and intuitive reasoning on the inner VNF workings, we use a piecewise model where two curves are fitted to the samples of each profiled configuration.
We define following analytic relations to model the performance:

\begin{itemize}[leftmargin=*]
\color{black}
\item In the non-saturated region, we choose an exponential function because of its characteristics similar to the observed trends: the function stays low in the beginning and only starts to rise rapidly later, as a transition phase to the saturated region.
\item The functions in the saturated region are based on intuitive assumptions of the internal VNF processing (see Section \ref{measurement_parameters}). The included parameters allow extra freedom to fit the slope and the x-axis intercept to the samples of each configuration. 
\end{itemize} 

\noindent
For the packet loss of the forwarding VNFs, we define an exponential curve which starts at zero and stays very low, until it starts to rise near a value $b$. After resource saturation, we model the packet loss by: 
\[
\mbox{packet loss (\%)} = \frac{A-P}{A} 100 = 100 (1-\frac{P}{A})
\]
where:\\
$A$ = Actual incoming packet rate. \\
$P$ = Processed packet rate (max throughput) at saturation.\\
The above equation shows that after saturation, the packet loss is a function inversely proportional with the incoming packet rate.  We use this information to define following analytic model for the packet loss ($x$ is the ingress packet rate):
  
\begin{equation}
\label{packet_fwd}
\begin{cases} 
-exp(-ab) + exp\left[a(x-b)\right], & \mbox{non-saturated } \\
100 (1 - \frac{c}{x-d}), & \mbox{saturated, with x>c+d}
\end{cases}
\end{equation}

For the cache server, we use the same exponential curve in the non-saturated region, but now the resulting value can be larger than zero at low request rates. After resource saturation, we model that the response time is given by:
\[
\mbox{response time (s)} = \frac{FS}{BW} U  
\]
where:\\
$BW$ = The maximum reachable download bandwidth \\
$FS$ = The average filesize of one file request\\
$U$ = The number of ongoing file requests\\
{\color{black}The actual value of BW is determined by both the processing time needed to prepare the request response and the link capacity to send the response. We assume that BW is constant (per configuration) after saturation, but we can only indirectly quantify its value by profiling.}
The above equation shows that after saturation, the response time is linearly proportional to the number of concurrent ongoing requests. We use this information to define following analytic model for the cache response time ($x$ is the number of concurrent ongoing requests):

\begin{equation}
\label{vnf_reqs}
\begin{cases} 
a + exp\left[b(x-c)\right], & \mbox{non-saturated } \\
d(x-e), & \mbox{saturated, with } x > e
\end{cases}
\end{equation}

As seen in Fig. \ref{curvefit_detail} we also need a method to split the metrics in a  non-saturated and saturated subset. Each sample subset is then fitted to its respective analytic model given in Eq. \ref{packet_fwd} and \ref{vnf_reqs}, by deriving optimal values for the parameters $a,b,c,d,e$.
In Algorithm \ref{algocf} we describe the procedure of fitting these curves to the profiled dataset.

\begin{algorithm}[h]
\label{algocf}
\color{black}
\SetAlgoLined
\KwData{Profiled VNF dataset}
\KwResult{$P$ = (curve parameters | profiled configurations)}
 $P \leftarrow$ empty VNF profile\;
 split the profiled dataset per unique configuration\;
 \For{each profiled configuration}{
  $S_{non-saturated} \leftarrow$ empty dataset\;
  $S_{saturated} \leftarrow$ empty dataset\;
  order the samples by increasing packet rate\;
  \While{sliding window over n samples}{
    standardize the samples in the sliding window\;
    $CoV_{res}\leftarrow covariance(CPU, packet\_rate)$\; 
    $CoV_{perf}\leftarrow  covariance(packet\_loss, packet\_rate)$\; 
    \If{$CoV_{perf}$ > $CoV_{res}$}{
      $S_{saturated}$.append(remaining samples)\;
      stop while loop\;
    }
    $S_{non-saturated}$.append(current window samples)\;
    move sliding window by 1 sample\;
  }
  curve fit $S_{non-saturated}$\;
  curve fit $S_{saturated}$\;
  calculate intersection/closest point\;
  $P$.append([fitted parameters, intersection point, configuration parameters])\;
 }
 \caption{Curve Fit training algorithm}
\end{algorithm}

{\color{black}
\noindent
Lines 4-14 describe how we split the samples in (non)saturated regions. 
For each plot in Fig. \ref{curvefit_detail} we calculate the covariance between the workload (x-axis metric) and each of the two y-axis metrics, in a sliding window. If the covariance with the performance (right y-axis metric) is greater, we enter the saturated (red) area.
This can also be visually examined, looking at Fig. \ref{curvefit_detail}:

\begin{itemize}[leftmargin=*]

\item In the \textit{non-saturated region} the resource metrics (CPU) vary more than the performance metrics (blue samples). 
The blue line shows the fitted function in this non-saturated region.

\item In the \textit{saturated region} the performance metrics (packet loss and response time, red samples) show the most variation. 
The CPU usage is saturated and remains more stable. 
The red line shows the fitted function in this saturated region.

\end{itemize}
}

The resulting VNF profile $P$ has a row for each profiled configuration (one configuration is a unique combination of resource allocation and workload settings).
For the packet forwarding VNFs (router, firewall and OVS) each configuration  is specified by (vCPU allocation, packetsize and number of flows).
Each configuration stores the fitted curve parameters ($a,b,c,d,e$ in Eq. \ref{packet_fwd} and \ref{vnf_reqs}) and the intersection point which indicates the boundary where the (non)saturated curve should be used.

For the Cache VNF, the metrics $packet\_loss$ and $packet\_rate$ in Algorithm \ref{algocf} are replaced by $response\_time$ and $concurrent\_requests$. 
Each Cache VNF configuration  is specified by (vCPU allocation, bandwidth allocation, filesize and cache hit ratio). For the other tested VNFs, each configuration is specified by (vCPU allocation, packetsize and number of flows).

\medskip
In order to predict the performance of an untrained configuration, we first lookup the surrounding configurations in the profiled dataset. Then we use the same method as used in the Interpolation model (see \ref{model_comparison}) to interpolate between the profiled fitted curves. This is exemplified in Fig. \ref{curve_fit_eg} (dashed line), where a monotonic and smooth curve results from the interpolation.

\medskip
We implemented this method using the Python SciPy library \cite{scipy}, which enables us to test and cross-validate this method in the same way as done with existing modelling methods such as kNN, ANN, etc.

\subsection{Accuracy of the Used Models}
\label{accuracy}
We have reported different accuracy measurements for the used methods in Table \ref{tab:accuracy}:

\begin{itemize}[leftmargin=*]
\item \textbf{r2}: R-squared is a statistical measure of how close the data are to the fitted model. For a multivariate model it is calculated as:  Explained variation / Total variation. Most of the models report a value near 100\% which indicates that the models explain most of the variability around their mean. Seeing the low deltas between the reported r2's, we conclude that this is not a significant score to compare the accuracy of the different models.

\item \textbf{MAE}: Mean Average Error. This is the mean value of the fitted residual errors.

\item \textbf{MAD}: Median Average Deviation. This is the median value of the fitted residual errors.

\item \textbf{RMSE}: Root Mean Squared Error. When the residuals of the fitted model are normally distributed, the RMSE depicts the standard deviation of the residuals. However, the difference between the MAE and MAD for most of the methods, seems to suggest that the distribution of the residual errors is skewed and larger errors are occurring at larger performance values.

\end{itemize}

\begin{figure}[!h]
\centering
\includegraphics[clip=true, trim=70 280 120 270, width=0.45\textwidth]{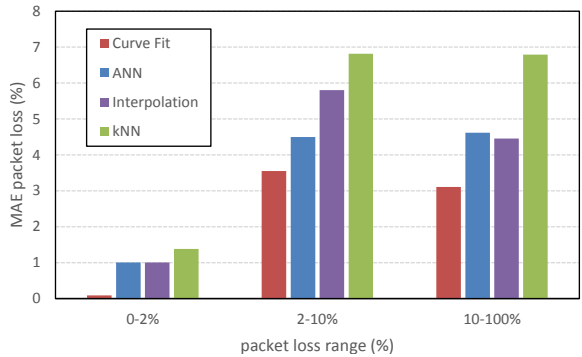}
\caption{Mean Absolute Error comparison for different methods and loss ranges, averaged over the packet forwarding VNFs.}
\label{MAE_loss}
\vspace{-3mm}
\end{figure}

In Fig. \ref{MAE_loss} we plot the MAE of the different methods, comparing only the performance (packet loss). We calculate the MAE in different buckets of the measured loss, averaged over all the packet forwarding VNFs (router, firewall and OVS). This indeed shows that larger errors are occurring at higher loss values, as suggested by the MAE  and MAD values.
The Curve Fit method produces the smallest errors.

\begin{figure}[!h]
\centering
\includegraphics[clip=true, trim=70 270 120 260, width=0.45\textwidth]{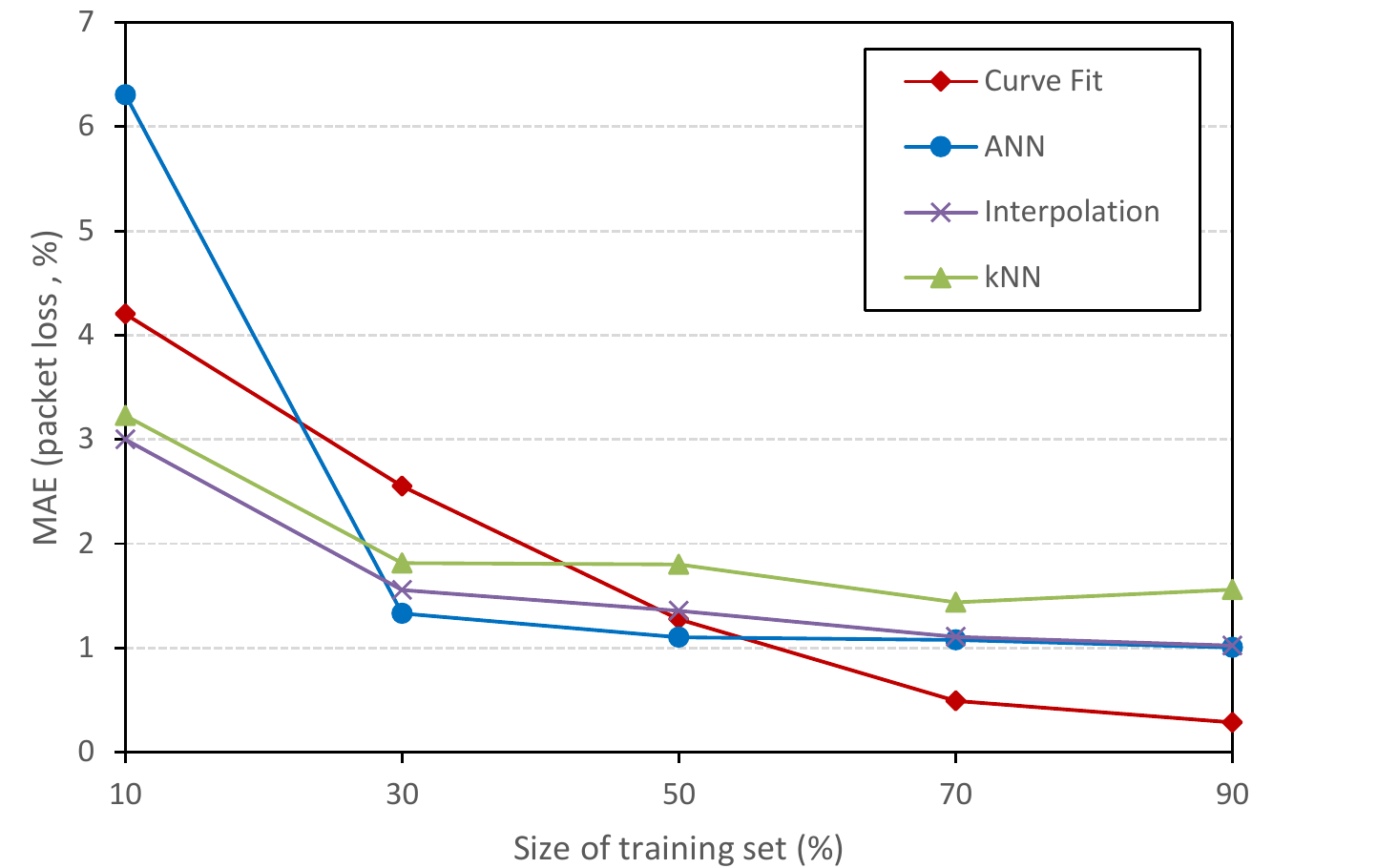}
\caption{Mean Absolute Error (in range 0-2\% packet loss) with increasing training set size.}
\label{MAE_general}
\vspace{-3mm}
\end{figure}

To have an even better understanding of the accuracy of the different methods, 
 we investigate the effect of decreasing the size of the training set in Fig. \ref{MAE_general}. We measure again the MAE in the 0-2\% packet loss bucket, using 5-fold cross-validation, but now we stepwise decrease the number of configurations used in the training set. At the right side of the graph, when using large enough training sets, we see indeed that the Curve Fit method has the smallest accuracy, which is in line with fig. \ref{MAE_loss} and Table \ref{tab:accuracy}.
At smaller training sets however, the Curve Fit method shows a decreasing accuracy, becoming worse than the other methods. The Curve Fit method appears to be the most sensitive to the size of the training set. This emphasizes the importance of a large enough dataset for training, and carefully controlling the boundaries in which the trained model is valid.

\subsection{Resource Allocation Recommendation}
\label{recommendation}
\begin{figure*}[!t]
\centering
\subfloat[OVS]{\includegraphics[width=.45\textwidth]{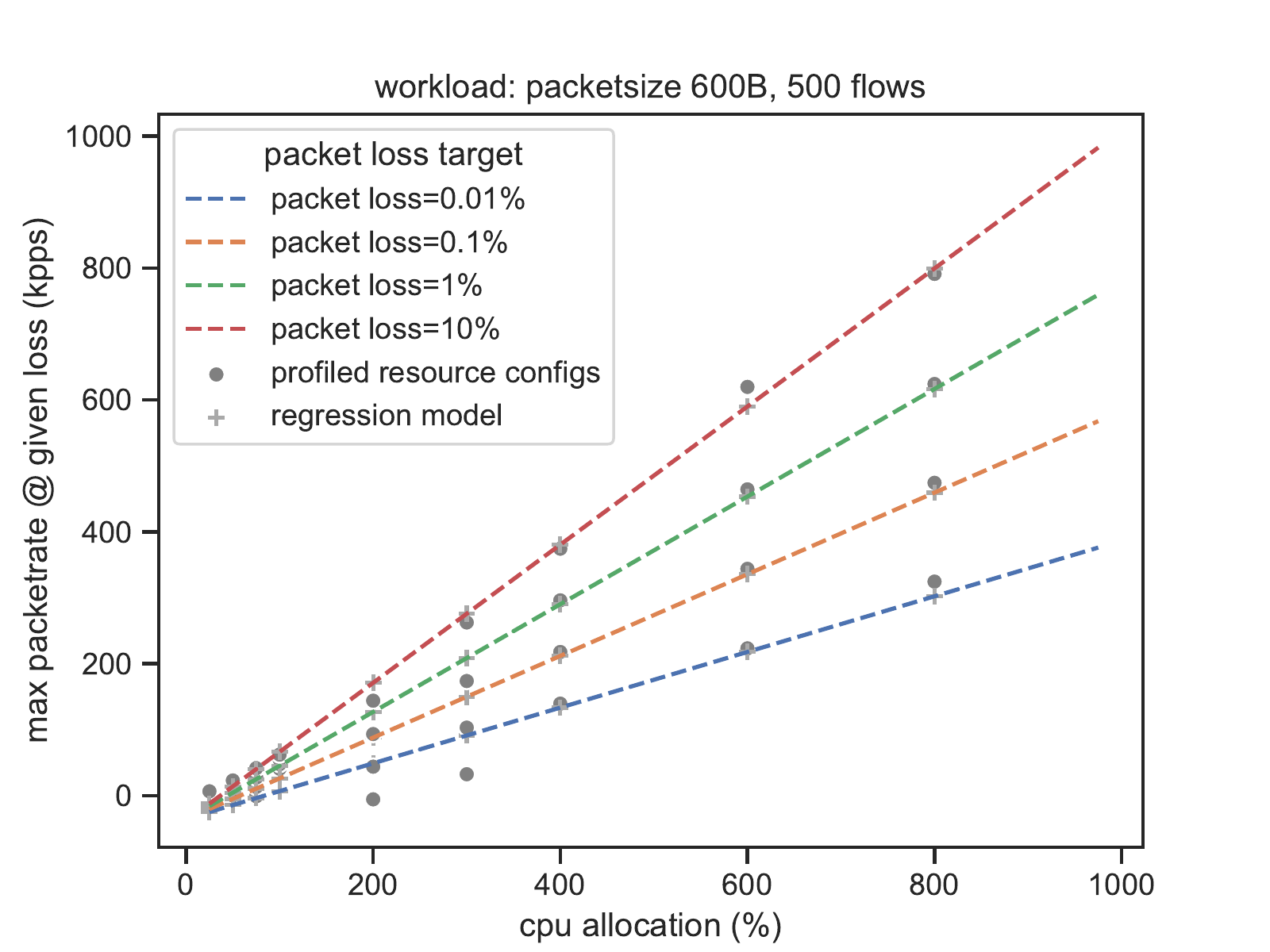}%
\label{ovs_recom}}
\hfil
\subfloat[Router]{\includegraphics[width=.45\textwidth]{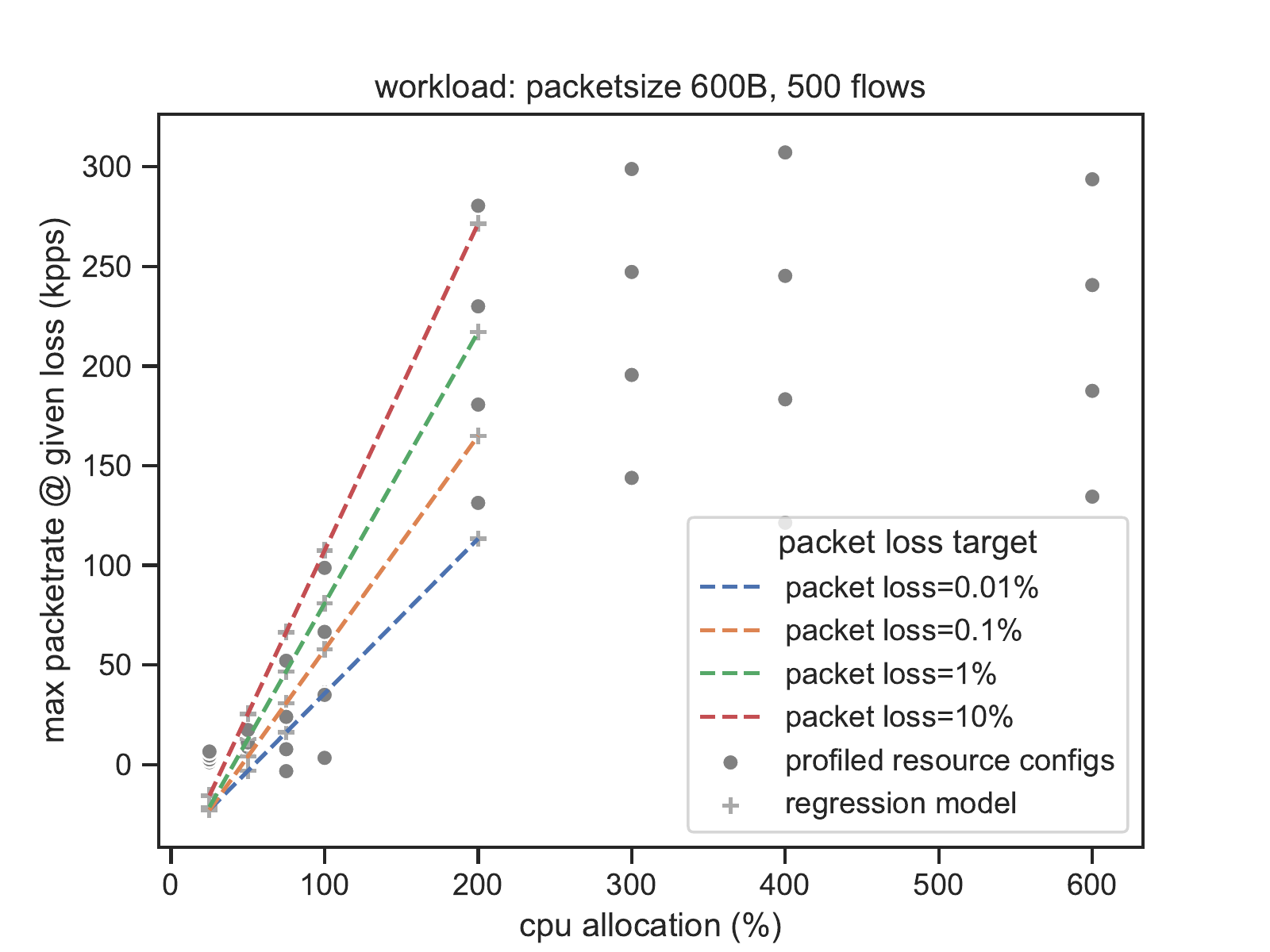}%
\label{router_recom}}
\hfil
\vspace{-4mm}
\subfloat[Firewall]{\includegraphics[width=.45\textwidth]{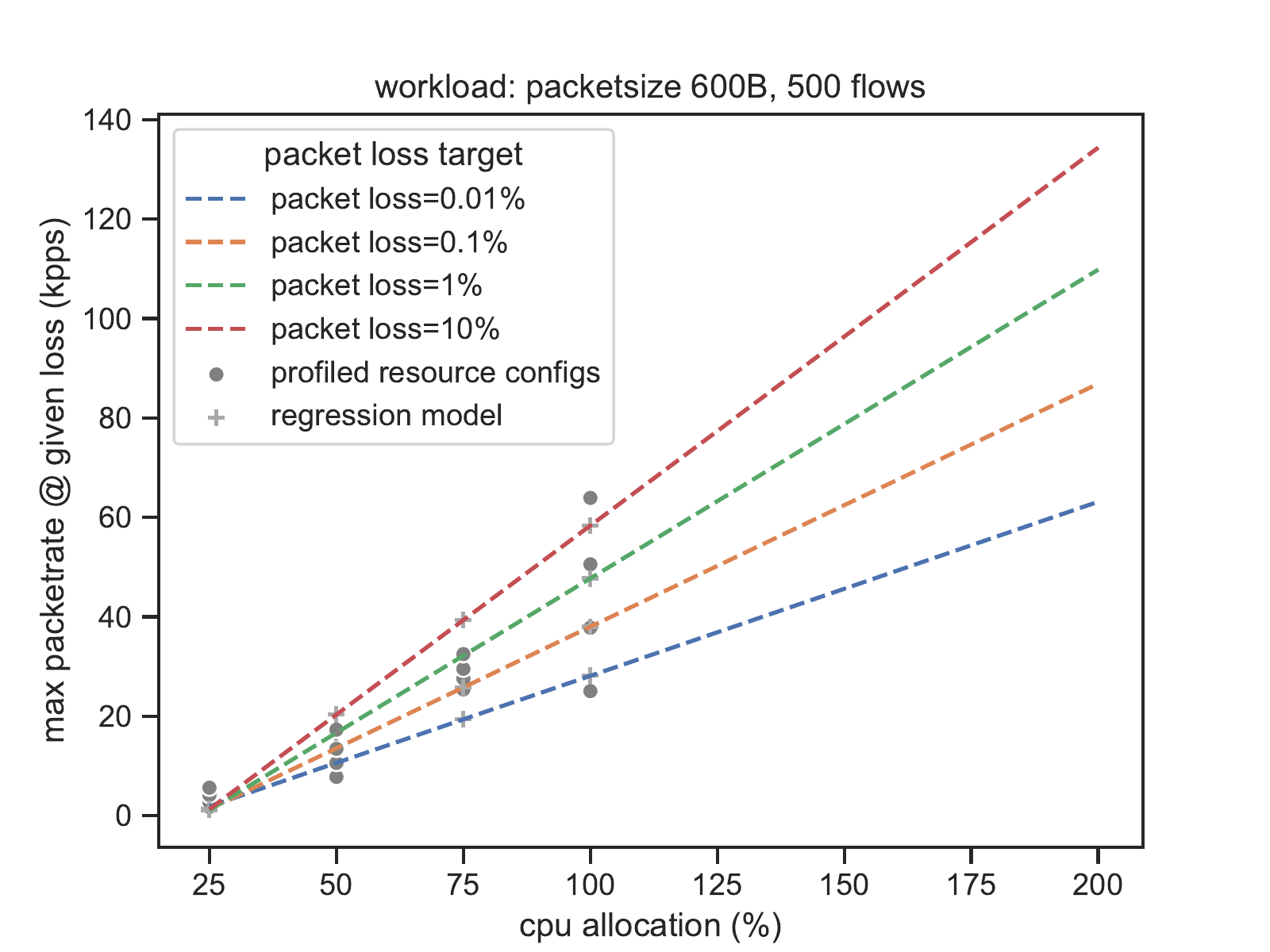}%
\label{fw_recom}}
\hfil
\subfloat[Cache]{\includegraphics[width=.45\textwidth]{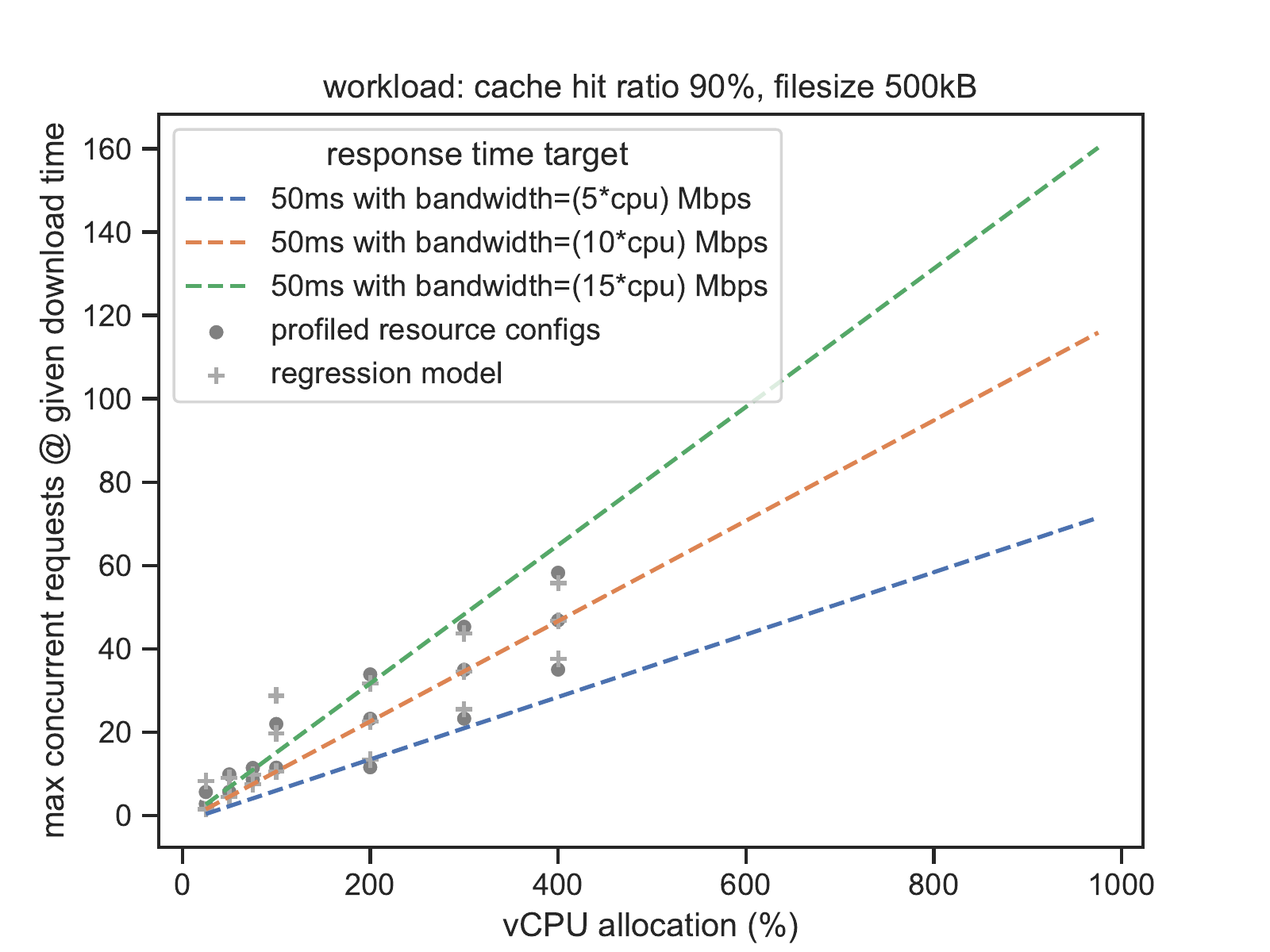}%
\label{squid_recom}}
\caption{{\color{black}VNF resource allocation recommendation. By regression, the performance trend of profiled resource configurations can be extrapolated.}}
\label{res_recommend}
\end{figure*}

In this section we describe how the obtained VNF profile can be used to give a recommendation for resource allocation. 
We therefore look back at Eq. \ref{fmodel2}: $f(wl, res)$ is modelled by the Curve Fit model described earlier. 
{\color{black}
We then use $f(wl, res)$ to calculate the maximum workload the VNF can process at each profiled resource allocation, within specified workload settings and KPI limits.
Next, we can interpolate and extrapolate this new dataset and predict the maximum workload the VNF can process for other resource allocations outside of the profiled set.
This procedure is given in algorithm \ref{algo_recom}:

\begin{algorithm}[h]
\label{algo_recom}
{\color{black}
\SetAlgoLined
\KwData{VNF Profile, $Perf_{target}$, $Wl_{target}$}
\KwResult{Recommended resource allocation}
$P_{res} \leftarrow$ empty dataset \;
\For{each profiled resource allocation $res$}{    
   Find the profiled workloads surrounding $Wl_{target}$\;
   $P \leftarrow$ empty dataset \;
   \For{each found profiled workload}{
    Lookup using the VNF profile model:\\
    $Wl_{max} \leftarrow$ $wl$ where $f(wl, res) == Perf_{target}$\;
    $P$.append([$Wl_{max}$, configuration parameters])\;
   }
   $Wl_{max}|res \leftarrow$ interpolate in $P$ the requested $Wl_{target}$ configuration\;
   $P_{res}$.append([$Wl_{max}|res$, configuration parameters])\;
    }
   /* $P_{res}$ now contains a predicted maximum workload for each profiled resource allocation */\\
   filter out resource allocations which bring no improvement\;
   train a regression model: $f(res)=Wl_{max}$  on $P_{res}$ to extrapolate $Wl_{max}$ to untrained resource allocations\;
   recommendation $\leftarrow$ find minimal $res$ where $f(Wl_{target}, res) \leq  Perf_{target}$\;

 \caption{Resource recommendation algorithm}

 }
\end{algorithm}

\begin{itemize}[leftmargin=*]
\item Line 3 uses a modified nearest neighbour algorithm, based on Euclidean distance, to  find and prioritize near configuration parameters at \textit{both} sides (i.e. with larger \textit{and}  with smaller parameters than $Wl_{target}$). This helps the interpolation method used further in the algorithm.
 
\item Line 6 uses the trained Curve Fit model to predict the VNF performance as explained in Section \ref{curve_fit_method}. However, to find $Wl_{max}$ we must inverse the functions in Eq. \ref{packet_fwd} and \ref{vnf_reqs}.

\item Line 10 represents again the Interpolation method as explained in  \ref{model_comparison}.

\item At the end, from line 13 onwards, we have a reduced dataset $P_{res}$, filtered by our given performance and workload targets. A simple regression can be used to model this dataset, since we observe a linear trend between the amount of allocated resources and the maximum workload (see Fig. \ref{res_recommend}). This corresponds to our intuitive understanding of a VNF implementation, where more resources proportionally allow more processed packets or requests.

\item Line 15 is hence an Ordinary Least Squares regression model with the allocated resources ($res$) as input and the according maximum processable workload ($Wl_{max}|res$) as output.

\end{itemize}
}

\noindent
A more practical example might clarify this further:\\
For the forwarding-based VNFs (router, firewall, OVS) we must specify following parameters in the SLA:
\begin{itemize}[leftmargin=*]
\item {\color{black}workload} target: (packetsize, number of flows)
\item performance target: packet loss
\end{itemize}
Using the Curve Fit model we predict which packet rate is expected at the targeted packet loss and configuration (line 5 in algorithm \ref{algo_recom}).
{\color{black}
We predict thus the maximum packet rate the VNF can process (at the specified packet loss) for every profiled resource allocation (line 10 in algorithm \ref{algo_recom}).
}

\noindent
Similar, for the Cache VNF, we must specify following parameters for algorithm \ref{algo_recom}:
\begin{itemize}[leftmargin=*]
\item {\color{black}workload} target: (filesize, cache hit ratio)
\item performance target: response time of cached file requests
\end{itemize}

\noindent
In Fig. \ref{res_recommend}, we show the predicted workload (y-axis, $Wl_{max}|res$) for a certain performance target in each profiled resource allocation (x-axis, $res$).
To get a recommendation for non-profiled resource allocations, we must interpolate or extrapolate between the obtained sample points. 
which is done by the regression model in Algorithm \ref{algo_recom}.
This is the function which is used by the service operator to determine an optimal resource allocation, meeting the SLA specifications. Using the graphs depicted in Fig. \ref{res_recommend}, a resource allocation can be chosen to process the targeted packetrate or concurrent incoming requests, constrained by a given packet loss or response time.

There is one special case illustrated by the router VNF in Fig. \ref{router_recom}.
This VNF does not show  further improvement when more than 2 vCPU are allocated. This should be detected, as no resource recommendation should be given beyond 2vCPUs for this router VNF (see also line 14 in Algorithm \ref{algo_recom}).

\medskip
As a side note we can report that the lookup time of all the samples of a VNF in Fig. \ref{res_recommend} is in the order of 100ms. This means that a delay of approx. 100ms is introduced in the orchestration procedure for recommending an adequate resource allocation. Fast booting network functions (e.g. implemented as unikernels) are however reported to boot in the order of 10ms. In this case the resource recommendation lookup will be a bottleneck. A possible way to mitigate this is to implement the network function in such a way that resource allocations can be dynamically adjusted, without requiring a reboot. Another possibility is that pre-trained regression models are pro-actively made for a number of fixed performance and configuration targets.

\section{More Use-Cases for VNF Profiles}
\label{usecases}
In the previous sections we have explained how the VNF profile can be constructed. Next, we see three immediate benefits which VNF profiles can bring to the operational framework of a service orchestration platform:
\begin{itemize}[leftmargin=*]
\item At \textbf{initial orchestration of the VNF}, the profile helps to calculate a sensible recommendation for resource allocation. This is done in order to meet the performance specifications in the SLA as soon as possible after the VNF has been started, avoiding the need for additional scaling cycles.

\item \textbf{During operation} of the VNF, fluctuations in the workload can cause performance degradations.  If the (long-term) workload fluctuations can be predicted, the VNF profile could map this to a resource allocation at which the VNF is able to keep up with the SLA/performance specification. Elastic scaling methods can make use of the VNF profile to calculate optimal resource allocation updates under changing workloads.   

\item The obtained VNF profile can serve as a \textbf{baseline behavior model }for the VNF. The operational performance and resource usage trends of the VNF can be compared against the profile by a healing algorithm to detect any deviations.

\end{itemize}
   
In this paper we have focused on the first two cases. The usefulness of the VNF profile becomes clear when we need to find a mapping between targeted workload and allocated resources, thus also when the VNF needs to scale. 
Referring back to Fig. \ref{curvefit_cache}, we can observe the following:\\
Suppose we target a response time of 50ms. According to the profiled performance curve, the number of concurrent requests can still increase while the VNF is operating in saturated mode, before the targeted response time is reached.
If we would scale the VNF already at the point where the CPU is saturated, we would always keep a very large margin to the targeted performance limit, and continuously overprovision the VNF.

\begin{figure}[!h]
\centering
\includegraphics[clip=true, trim=0 15 0 20,width=0.5\textwidth]{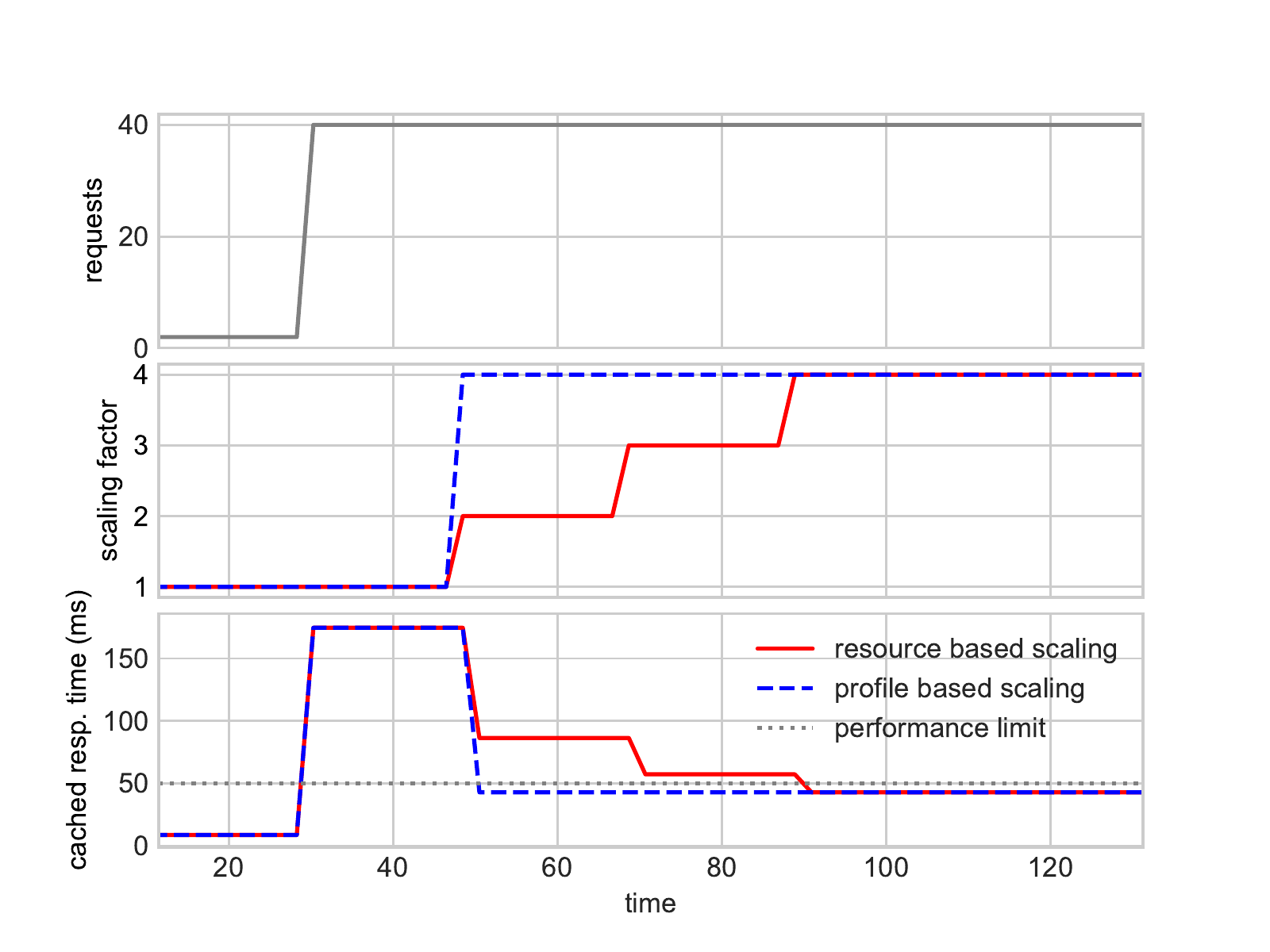}
\caption{At a workload surge, the profile can be used to determine the needed resources, resulting in less scaling iterations (less vCPU scaling steps in the middle plot).}
\label{fig:scaling}
\vspace{-3mm}
\end{figure}

In Fig. \ref{fig:scaling} we show how the VNF profile can improve the scaling efficiency. We take the Cache VNF as an example. The upper plot shows a sudden surge in the workload, the number of requests suddenly increases.
The middle plot shows how different scaling strategies would react to this. 
Resource-based scaling triggers have no idea how much resources to add to bring performance again to an acceptable level, therefore multiple iterations can happen.
Resource-based scaling algorithms aim to bring the VNF below a certain resource usage threshold, without any knowledge how the performance is related to the resource usage.  
Using the VNF profile, we can lookup which resource allocation meets the performance target for the number of requests at the top of the surge. Therefore only one single scaling iteration is executed. 
Profile-based scaling algorithms can more optimally estimate how much resources are required to meet a performance threshold.
This example shows also that the VNF profile can be used to pro-actively add more resources if pre-defined workload surges are expected.

\noindent
Regarding profile-based scaling, we can conclude the following:
\begin{itemize}[leftmargin=*]
\item Scaling triggers based on resource usages might be suboptimal, since also in saturated mode, the VNF performance can be good enough.
Default autoscale functions in Openstack or Kubernetes for example operate this way.
\item When using performance-based scaling triggers, we optimally make use of the allocated resources. In this case we can let VNF also operate in the saturated region.
\item Without a VNF profile we can never predict for sure if a certain resource allocation offers enough performance or not.
\end{itemize}

Another challenge is the performance fluctuation, inherent to the cloud's operational context. 
In a typical use-case, the vendor supplies the application in a portable virtualized form (like a virtual machine, container or unikernel). 
The operator and infrastructure provider can choose themselves how to optimally deploy the application.
This means that the service operator may implement different strategies regarding resource isolation (isolated or shared resource mapping), inducing noisy neighbour effects: 
To limit under-utilization and save power, datacenter operators can overcommit
their hardware by allowing multiple VNFs on the same CPUs.
This creates of course starvation if each VNF is requesting its complete allocated resource share \cite{skewness}.
Isolation between processes can also be compromised on a higher level, e.g. containers sharing the same kernel.
Similarly, the service provider can route multiple tenants to the same shared VNF, making it difficult to predict the total workload.

Several measurement campaigns gave insight in the performance fluctuations of multiple commercial cloud providers. Empirical results show significant performance differences for comparable instances on the public cloud \cite{muhammad2017transparent}. 
One of the most striking learnings from this research is that cloud-hosted VNF performance can suffer from short-lived but frequent episodes of very severe performance degradation. CPU, memory and especially IO-bound workloads can vary greatly (double-digit percentage fluctuations).
Other measurements also show that
multi-tenancy has a dramatic impact on performance and predictability\cite{leitner2016patterns}.

The VNF profile can be used to tackle these fluctuations.
It allows to translate any foreseen resource fluctuations to according performance fluctuations. If we take into account these predicted performance fluctuations, we can take the necessary counter actions to better guarantee the SLA.

\section{Conclusion and Summary}
\label{conclusion}

We have discussed the use-cases for VNF profiling and investigated which methods work best to model VNF performance.
Using four different VNF implementations we have quantified the accuracy of the investigated methods. Although widely used in various domains, our results show that regression, k-Nearest Neighbors, interoplation and Neural Networks do not offer good accuracy when modelling the typical performance trends of VNFs. 
{\color{black}
This is related to the non-linear relations between VNF parameters, noisy variation of the performance measurements and a low quantity of profiled data points due to time restrictions.
These drawbacks are mitigated in our experiments, since we use curve fitting to model the profiled data points and interpolate between the fitted curves to achieve the highest prediction accuracy of the VNF performance. 
}
Using this newly proposed modelling approach, it is shown how the modelled VNF profile can assist the service operator by providing resource recommendations for a VNF. This allows performance specifications in the SLA to be mapped to practical resource allocations. This is not only useful at initial orchestration of the VNF, but also when the VNF needs to scale to new resource allocations, due to changing workloads or dynamic cloud-native performance fluctuations.

\vspace{-5mm}
\smaller
\section*{Acknowledgment}
This work has been performed in the framework of the NGPaaS and 5GTANGO project, funded by the European Commission under the Horizon 2020 and 5G-PPP Phase2 programmes, resp. under Grant Agreement No. 761 557 and 761 493 (http://ngpaas.eu) (https://www.5gtango.eu). This work is partly funded by UGent BOF/GOA project 'Autonomic Networked Multimedia Systems'. 

\ifCLASSOPTIONcaptionsoff
  \newpage
\fi



\vspace{-3mm}
\bibliographystyle{IEEEtran}
\bibliography{refs}

\end{document}